\documentclass[aps,pre,onecolumn,showpacs,showkeys,a4paper]{revtex4-1}

\usepackage{stmaryrd}

\usepackage[utf8]{inputenc}

\usepackage{graphicx} 
\usepackage{amsmath}
\usepackage{amsfonts}
\usepackage{amscd}
\usepackage{amssymb}
\usepackage{fullpage}
\usepackage{color}

\usepackage{float}
\usepackage{empheq} 
\usepackage{fancybox}

\usepackage{epsf}
\usepackage{epsfig}
\usepackage{epstopdf}

\usepackage{hyperref}
\hypersetup{
    colorlinks,
    citecolor=blue,
    filecolor=black,
    linkcolor=blue,
    urlcolor=black
}

\begin{document}

\title{Generic Dynamical Phase Transition \\ in One-Dimensional Bulk-Driven Lattice Gases with Exclusion}
\author{Alexandre Lazarescu}
\affiliation{Complex Systems and Statistical Mechanics Group, University of Luxembourg}
\begin{abstract}
Dynamical phase transitions are crucial features of the fluctuations of statistical systems, corresponding to boundaries between qualitatively different mechanisms of maintaining unlikely values of dynamical observables over long periods of time. They manifest themselves in the form of non-analyticities in the large deviation function of those observables.

In this paper, we look at bulk-driven exclusion processes with open boundaries. It is known that the standard asymmetric simple exclusion process exhibits a dynamical phase transition in the large deviations of the current of particles flowing through it. That phase transition has been described thanks to specific calculation methods relying on the model being exactly solvable, but more general methods have also been used to describe the extreme large deviations of that current, far from the phase transition. 

We extend those methods to a large class of models based on the ASEP, where we add arbitrary spatial inhomogeneities in the rates and short-range potentials between the particles. We show that, as for the regular ASEP, the large deviation function of the current scales differently with the size of the system if one considers very high or very low currents, pointing to the existence of a dynamical phase transition between those two regimes: high current large deviations are extensive in the system size, and the typical states associated to them are Coulomb gases, which are highly correlated ; low current large deviations do not depend on the system size, and the typical states associated to them are anti-shocks, consistently with a hydrodynamic behaviour. Finally, we illustrate our results numerically on a simple example, and we interpret the transition in terms of the current pushing beyond its maximal hydrodynamic value, as well as relate it to the appearance of Tracy-Widom distributions in the relaxation statistics of such models.
\end{abstract}

\maketitle

\tableofcontents

\newpage

\section{Introduction}
\label{I}

The main mission of statistical physics is to bridge the gap between different levels of description of physical systems, from a microscopic level where we know the laws governing each of the myriad of components of the system and their interactions, to a macroscopic level where the whole system is described through emergent laws relating only a handful of global observables. The law of ideal gases is an extreme example of that, where three quantities are enough to describe the typical state of a somewhat caricatural system in a rather specific set-up, but it is essentially the same question that is asked when one want to describe, for instance, a model of particles moving and interacting on a lattice, by a Langevin equation on just the local average density of particles with a conserved Gaussian noise. Obtaining such a hydrodynamic description, and determining whether the Gaussian noise does reproduce faithfully the fluctuations of the lattice model at that scale, is in general an extremely challenging problem.

For systems that are close to equilibrium, where detailed balance is broken only infinitesimally in the large size limit, such as boundary-driven or weakly bulk-driven models, the aforementioned macroscopic hydrodynamic description is known to hold quite generally, under the names of macroscopic fluctuation theory (MFT, \cite{Bertini2007}) and additivity principle \cite{PhysRevLett.92.180601}. The question becomes more interesting when one considers systems that are far from equilibrium, where no such general result exists. There, from an analytical perspective, one is mostly restricted to study specific models or classes of models which are amenable to calculations, which often means exactly solvable, such as zero-range processes \cite{Harris2005,Chleboun2016} or exclusion processes \cite{Derrida199865,0034-4885-74-11-116601}. Moreover, since the question involves not only the typical behaviour of the systems at a macroscopic scale, but their fluctuations as well, the relevant framework is that of large deviation functions \cite{Touchette20091}, which are the dynamical equivalent of equilibrium free energies.

The most interesting feature of those large deviation functions are so-called \textit{dynamical phase transitions}. Just like their equilibrium equivalents, they appear as non-analyticities which correspond to boundaries between qualitatively different behaviours of the system when changing its parameters. Unlike for equilibrium phase transitions, however, dynamical control parameters are usually abstract quantities which do not correspond to actual tunable parameters of the physical model \cite{Jack2015a}, and those different qualitative behaviours correspond to the best way in which the system can fluctuate to maintain an atypical value of some observable for a long time, rather than stable state changes that could be observed experimentally by tuning the environment. This is not to say that dynamical phase transitions have no experimentally measurable consequences, as for instance models which are critical in that respect (i.e. where the stationary state sits precisely at a dynamical phase transition) will show non-Gaussian fluctuations, anomalous diffusive behaviours and special dynamical exponents, a famous example being models in the KPZ universality class \cite{Spohn2016,Corwin2016}. Identifying those dynamical phase transitions is quite crucial when attempting to describe the macroscopic behaviour of stochastic systems: they define the domains of applicability of any proposed description.

~~

In this paper, we focus on one-dimensional bulk-driven open simple exclusion processes (SEP), where particles jump from site to neighbouring site on a one dimensional lattice, with a bias towards one side (e.g. the right), and can enter or exit the system only at the boundaries. The \textit{simple exclusion} constraint means that the particles interact through hard-core repulsion: at most one particle can be on a given site at a given time. These models are among the most studied in non-equilibrium statistical physics, both analytically for the exactly solvable versions \cite{derrida1993exact,1751-8121-40-46-R01,Lazarescu2015} (which can also have periodic boundary conditions \cite{derrida1998exact,Prolhac2009} and symmetric or weakly asymmetric jumps \cite{Enaud2004,Prolhac2009a}), and numerically in the case of non-solvable generalisations \cite{1742-5468-2008-06-P06009,Greulich2008,Ciandrini2010}.

For such transport models, the most important observables are usually the macroscopic currents of particles (or whatever charges are being transported), which are a direct consequence of the system being driven out of equilibrium, and which can be expected to play an important role in its macroscopic behaviour. In particular, any dynamical phase transition that the system might undergo will most likely be visible in the large deviation function of those currents, which is a contraction of the full joint large deviation function of currents and densities. Indeed, such transitions have been found, for instance in the weakly asymmetric version of the model, both for periodic boundary conditions \cite{PhysRevE.72.066110,Bodineau2008,appert2008universal,Simon2011,Espigares2013} and open ones \cite{Lecomte2010,Baek2016a}.

A dynamical phase transition has also been found in the large deviations of the current of the \textit{asymmetric} simple exclusion process, both for total asymmetry (TASEP, where particles jump only to the right) and partial asymmetry (PASEP or simply ASEP, where particles can jump backwards as well), and both in the periodic and the open geometry. Those models are integrable (in the sense of quantum integrability \cite{Faddeev1996,Lazarescu2014}, through its connexion with the XXZ spin chain \cite{sandow1994partially}), meaning that one can in principle obtain an exact expression for the large deviation function, and describe the transition analytically. This was first done in a periodic and totally asymmetric setting in \cite{derrida1998exact} and \cite{derrida1999universal}, which was then extended to the partially asymmetric case \cite{Prolhac2009,prolhac2010tree}. In the open setting, part of the large deviation function was correctly conjectured in \cite{Bodineau2006} and later confirmed in \cite{DeGier2011a}, and the full large deviation function for the TASEP was obtained soon afterwards \cite{Lazarescu2011} and later extended to the ASEP \cite{gorissen2012exact,Lazarescu2014}. In all of these works, the results were obtained using integrability methods such as the Bethe Ansatz, and are in principle restricted to small fluctuations of the current because of approximations made towards the large size limit, except for \cite{Popkov2010} which deals with the limit of very high currents in the periodic ASEP. 

Finally, the complete dynamical phase diagram of the current for the open ASEP was obtained in \cite{Lazarescu2013,Lazarescu2015}, by combining the aforementioned results for small fluctuations with exact diagonalisation methods for extreme values of the current. It was found that in the very high current limit, the system effectively behaves like a discrete Coulomb gas, as it does in the periodic case \cite{Popkov2010}, with a large deviation function proportional to the size $L$ of the system ; in the very low current limit, the system has an effective dynamics involving only anti-shock states, and the large deviation function does not depend on $L$. The whole phase diagram is then obtained by interpolating between those tree regimes, and by conjecturing that the MFT can be used to obtain a large part of it (which is compatible with all the exact results). A dynamical phase transition is identified, which includes the stationary maximal current phase, and separates the two different scalings in $L$.

~

As we mentioned, a large part of those results are obtained by integrability methods, which are unfortunately entirely specific to that very special model. Luckily, the aforementioned exact diagonalisation methods for extreme currents do not require the model to be integrable, which means that they can be applied in more general settings. Indeed, in this paper, we consider a much broader class of models, by adding two features to the ASEP: arbitrary spatial inhomogeneities in the jump rates, and an arbitrary short range interaction between the particles. We will see that the same methods apply (albeit in a much more involved way) and yield essentially the same results: the large deviation function of the current is proportional to $L$ for high currents, and independent of $L$ for low currents. This points to the existence of a dynamical phase transition separating those two regimes, although this time no description of the transition itself is available, and in particular no information on the critical exponents associated to it, which leads us to qualify the transition as \textit{generic} rather than \textit{universal}.

The structure of this paper is as follows. In section \ref{II}, we first introduce the reader to the models and formalism we will be using throughout the paper, along with a few useful standard results. In section \ref{III}, we focus on the high current limit, where we first recall the results pertaining to the regular ASEP, before extending them rather straightforwardly to our generalised versions. In section \ref{IV}, which contains the main technical result of this paper, we do the same for the low current limit, by finding bounds on the behaviour of both the large deviation function of the current, and the shape of the typical states associated to those fluctuations. We then provide numerical illustrations for those results in section \ref{V}, as well as an interpretation of the origin of the dynamical phase transition in terms of maximal hydrodynamic currents, and discuss its relation to the KPZ universality class \cite{BenArous2011} and third order phase transitions \cite{LeDoussal2016}. We finally conclude and discuss a few extensions that our result could receive in the future.

\newpage

\section{Models and formalism}
\label{II}

In this first section, we define the models that we intend to study, and present the formalism in which they will be treated, as well as the mathematical tools we will use in order to access the fluctuations of the current of particles flowing through them.

\subsection{Markov matrix and master equation}
\label{IIa}

\begin{figure}[ht]
\begin{center}
 \includegraphics[width=0.8\textwidth]{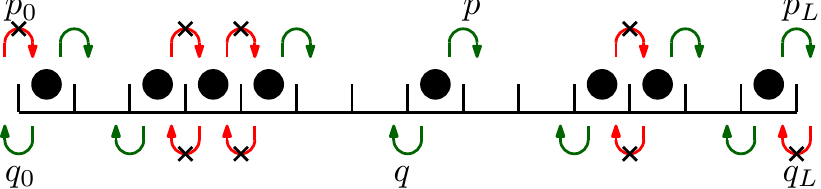}
  \caption{Dynamical rules for the ASEP  with open boundaries. The jumps shown in green are allowed by the exclusion constraint. Those shown in red and crossed out are forbidden.}
\label{fig1}
 \end{center}
 \end{figure}

The basic model upon which we aim to build is the asymmetric simple exclusion process (ASEP). It is a Markov process in continuous time defined on a finite one-dimensional lattice of size $L$, be it periodic or open and connected to reservoirs at both ends. Each site can be either empty, or holding one particle, and particles hop from site to site with a rate $p$ to the right and $q<p$ to the left (as if driven by a field $\log(p/q)$). In the open case, particles can enter the system on the first site with rate $p_0$ and on the last with rate $q_L$, and exit from the first site with rate $q_0$ and from the last with rate $p_L$. In all these cases, if the target site is already occupied, the exclusion rule means the particle cannot jump. The rates for the open case are summarised on fig.\ref{fig1}.

We will also be considering a simpler version called the totally asymmetric simple exclusion process (TASEP), where the particles can only jump to the right, which is to say that $q_0=q=q_L=0$. Note that we will be focusing on the open case, but that everything we will say is easily transposed to the periodic case, as will be regularly pointed out.

~

To put that into equations: states of the system are written as configurations ${\cal C}=\{\tau_i\}$, where $\tau_i=0$ if site $i$ is empty and $\tau_i=1$ if it is occupied. The probability vector $|P_{t}\rangle$, of which an entry $P_{t}({\cal C})$ gives the probability to be in configuration ${\cal C}$ at time $t$, obeys the master equation
\begin{equation}\label{MP}
\frac{d}{dt}|P_{t}\rangle=M|P_{t}\rangle
\end{equation}
where $M$ is the Markov matrix of the open ASEP:
\begin{equation}\label{M}
M=m_0+\sum\limits_{i=1}^{L-1}M_i +m_L
\end{equation}
with
\begin{equation}\label{M2}
m_0=\begin{bmatrix} -p_0 & q_0 \\ p_0 & -q_0  \end{bmatrix}~,~ M_{i}=\begin{bmatrix} 0 & 0 & 0 & 0 \\ 0 & -q & p & 0 \\ 0 & q & -p & 0 \\ 0 & 0 & 0 & 0 \end{bmatrix}~,~m_L=\begin{bmatrix} -q_L & p_L \\ q_L & -q_L  \end{bmatrix}.
\end{equation}
It is implied that $m_0$ acts as written on site $1$ (and is represented in basis $\{0,1\}$ for the occupancy of the first site), and as the identity on all the other sites. Likewise, $m_L$ acts as written on site $L$, and $M_i$ on sites $i$ and $i+1$ (and is represented in basis $\{00,01,10,11\}$ for the occupancy of those two sites). Each of the non-diagonal entries represents a transition between two configurations that are one particle jump away from each other. Note that we will be using the convention where $w(\cal{C},\cal{C}')$ is a transition from $\cal{C}'$ to $\cal{C}$, consistently with $M$ acting to the right on $|P_{t}\rangle$.

~

We will be generalising this model in two ways. First, by adding an interaction potential $V({\cal C})$ to configurations, which does not have to be two-body or translation-invariant (note that we will later assume $V$ to be short-range and bounded in order to prove the results in section \ref{IV}). This is done by multiplying the transition rate from ${\cal C}'$ to ${\cal C}$ by ${\rm e}^{(V({\cal C}')-V({\cal C}))/2}$. If the system were in equilibrium (if for instance $p=q$ in the periodic case), the stationary probability of ${\cal C}$ would then be ${\rm e}^{-V({\cal C})}$ up to a normalisation. Note that in order not to introduce too many irrelevant quantities, the Boltzmann inverse temperature is not written explicitly (i.e. either taken equal to $1$, or as an inverse unit of energy).

Secondly, by adding on-site inhomogeneities in the jump rates, i.e. by having them depend on space: $p$ and $q$ become $p_i$ and $q_i$, where $i$ is the label of the bond involved in the transition, starting at $1$ between the first and second site (so that a particle on site $i$ jumps to the right with rate $p_i$ and to the left with rate $q_{i-1}$), consistently with the notation used for the boundary rates. This second generalisation merely adds an index $i$ to the entries of $M_i$, but the first one modifies the structure of $M$: the entries of $M_i$ can now depend on all the details of the initial and final configurations, so that they are not effectively of dimension $4$ any more. Note that part of the inhomogeneity can be absorbed in $V$, so that these two generalisations are not orthogonal, but this will not be important for us.

~

We will be generically writing the transition rates of our process as $w(\cal{C},\cal{C}')$, and we will take ${\cal C}\sim{\cal C}'$ to mean that there is a transition from ${\cal C}'$ to ${\cal C}$. It will be useful for future calculations to decompose the Markov matrix of this generalised simple exclusion process into three pieces:
\begin{itemize}
\item $M^+$, containing the rates for jumps to the right, of the form ${\rm e}^{-V({\cal C})/2}~p_i~{\rm e}^{V({\cal C}')/2}$ ;
\item $M^-$, containing the rates for jumps to the left, of the form ${\rm e}^{-V({\cal C})/2}~q_i~{\rm e}^{V({\cal C}')/2}$ ;
\item $M^d$, containing the escape rates, i.e. the diagonal part of $M$, of the form $-\sum_{{\cal C}\sim{\cal C}'}w(\cal{C},\cal{C}')$.
\end{itemize}
Note that the addition of the potential $V$ to $M^+$ and $M^-$ is a conjugation (division to the left, multiplication to the right) of the potential-less rates with a diagonal matrix ${\rm e}^{V/2}$, but this is not the case for $M^d$.

\subsection{Deformed Markov matrix for the current}
\label{IIb}

Now that we have defined our processes, we can see how to access the statistics of the stationary current. The simplest, most tractable way to do that, for our purposes, is through the cumulants of the current.

Let us say that we want to know, for instance, the stationary statistics of the time-averaged current of particles $j_0$ across the bond between the left reservoir and the first site. We should, in principle, from a given initial condition, look at all the possible histories of the system for a certain runtime $t$, count the number of particles crossing that bond in either direction, and compute the probability that the difference of these two numbers is close to $t j_0$. A simpler way to proceed is to introduce a fugacity (or counter or Lagrange multiplier) ${\rm e}^{\mu_0}$ for that algebraic number of ingoing jumps, and multiply, in $M$, the rate which increases that count by one (i.e. in $M^+$) by ${\rm e}^{\mu_0}$ and the rate which decreases it (i.e. in $M^-$) by ${\rm e}^{-\mu_0}$. By using these deformed rates instead of the original ones in our process (which is not a proper Markov process any more, since we have not modified $M^d$), every history will be multiplied by as many fugacities as there was jumps, i.e. by a factor ${\rm e}^{t j_0 \mu_0}$. Summing these weighted probabilities and taking derivatives with respect to $\mu_0$ will then yield the moments of $t j_0$, which we can then relate to their probability distribution.

~

\begin{figure}[ht]
\begin{center}
 \includegraphics[width=0.6\textwidth]{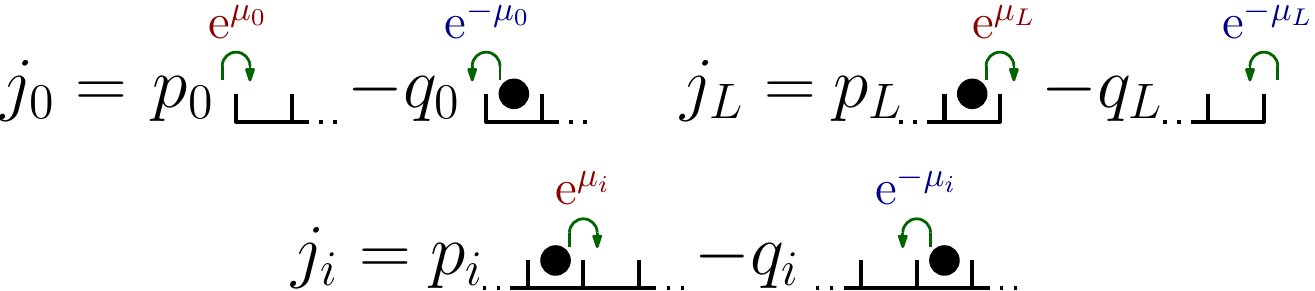}
  \caption{Transitions contributing to each of the physical currents $j_i$, along with the fugacities applied to each of the (non-diagonal) transition rates.}
\label{Currents}
 \end{center}
 \end{figure}

This can be rephrased neatly in a few equations. For the sake of compactness, we will consider the case where $V=0$, so that we can write $M$ in a simple way, but it is straightforward to include the potential as well. Let us now assume that we want to monitor all of the time-averaged currents $j_i$, across each of the $L+1$ bonds in the system, independently, in order to obtain their large deviation function $g(j)$ defined through
\begin{equation}
\mathbb{P}_{0..t}\bigl[\{[\#~i\rightarrow i+1]\}\sim \{tj_i\}\bigr]\approx \mathrm{e}^{-tg(j)}
\end{equation}
in the large $t$ limit, where $[\#~i\rightarrow i+1]$ is the algebraic number of jumps made between sites $i$ and $i+1$ from time $0$ till $t$. We associate a fugacity ${\rm e}^{\mu_i}$ to each current (fig.\ref{Currents}), and write simply $\mu$ for the vector containing the $\mu_i$'s.

Consider then the following matrix
\begin{equation}\label{Mmu}
M_{\mu}=m_0(\mu_0)+\sum_{i=1}^{L-1} M_{i}(\mu_i)+m_L(\mu_l)
\end{equation}
with
\begin{equation}\label{Mmui}
m_0(\mu_0)=\begin{bmatrix} -p_0 & q_0{\rm e}^{-\mu_0} \\ p_0{\rm e}^{\mu_0} & -q_0  \end{bmatrix}~,~ M_{i}(\mu_i)=\begin{bmatrix} 0 & 0 & 0 & 0 \\ 0 & -q_i & p_i{\rm e}^{\mu_i} & 0 \\ 0 & q_i{\rm e}^{-\mu_i}& -p_i & 0 \\ 0 & 0 & 0 & 0 \end{bmatrix}~,~m_L(\mu_L)=\begin{bmatrix} -q_L & p_L{\rm e}^{\mu_L} \\  q_L{\rm e}^{-\mu_L} & -p_L  \end{bmatrix}
\end{equation}
(where, as before, it is implied that $m_0$ acts as written on site $0$ in the basis $\{0,1\}$ and as the identity on the other sites, and the same goes for $m_L$ on site $L$; similarly, $M_i$ is expressed by its action on sites $i$ and $i\!+\!1$ in the basis $\{00,01,10,11\}$ and acts as the identity on the rest of the system).

~

As we said earlier, summing the weights of histories obtained by using this matrix as a generator for a time $t$ from an initial state $P_0$ yields the generating function of the joint moments of all the currents up to a trivial dependence on $t$. Taking the logarithm of that generating function and dividing by $t$ therefore gives the (rescaled) generating function of the cumulants of the current $E$, in the limit of large times:
\begin{equation}\label{Emui}
E(\mu)=\lim\limits_{t\rightarrow\infty}\frac{1}{t}\bigl( \langle 1|{\rm e}^{tM_{\mu}}|P_0\rangle \bigr)
\end{equation}

It is straightforward to show from that expression that $E(\mu)$ is in fact the largest eigenvalue of $M_{\mu}$: for $t$ large, we can write
\begin{equation}
{\rm e}^{tM_{\mu}}\sim {\rm e}^{t\Lambda(\mu)}|P_{\mu}\rangle \langle \tilde{P}_{\mu}|
\end{equation}
where $\Lambda(\mu)$ is the largest eigenvalue of $M_{\mu}$ and $|P_{\mu}\rangle$ and $ \langle \tilde{P}_{\mu}|$ are the corresponding eigenvectors. Injecting this in (\ref{Emui}) yields $E=\Lambda$. All of these elements hold important information regarding the stationary fluctuations of the current and the fluctuating states themselves. We can show, by making a saddle-point approximation for large $t$ in the definition of $E$, that \cite{Touchette20091,Lazarescu2015}:
\begin{itemize}
\item G\"artner-Ellis theorem: the large deviation function of the currents, $g(j)$, where $j$ is the vector holding all the individual currents $j_i$, is the Legendre transform of $E(\mu)$, i.e.
\begin{equation}\label{gEi}
g(j)=j.\mu-E(\mu)~~~~{\rm with}~~~~j_i=\frac{{\rm d}}{{\rm d}\mu_i} E(\mu).
\end{equation}
\item The right eigenvector holds probabilities of final configurations conditioned on the currents that have been observed, up to a normalisation, i.e.
\begin{equation}
P_\mu({\mathcal C})\propto{\rm P}\Bigl({\mathcal C}_t={\mathcal C}~\Big|~j_i\!=\!\frac{{\rm d}}{{\rm d}\mu_i} E(\mu)\Bigr).
\end{equation}
\item The left eigenvector holds probabilities of initial configurations conditioned on the currents that will be observed, up to a normalisation, i.e.
\begin{equation}
\tilde{P}_\mu({\mathcal C})\propto{\rm P}\Bigl({\mathcal C}_0={\mathcal C}~\Big|~j_i\!=\!\frac{{\rm d}}{{\rm d}\mu_i} E(\mu)\Bigr).
\end{equation}
\item The product of these two probabilities gives the stationary probability of configurations (far from the initial and final time) conditioned on the currents that are observed, up to a normalisation, i.e.
\begin{equation}\label{sensemble}
P_\mu({\mathcal C})\tilde{P}_\mu({\mathcal C})\propto{\rm P}\Bigl({\mathcal C}~\Big|~j_i\!=\!\frac{{\rm d}}{{\rm d}\mu_i} E(\mu)\Bigr).
\end{equation}
\end{itemize}

In all of these cases, the currents we consider are time-averaged, but for a finite-sized system, they also correspond to stationary instantaneous currents (in other words, a given average current is best realised, over a long period, by a constant instantaneous current of the same value). These properties are not specific to one-dimensional simple exclusion processes, but are in fact valid for any currents in any Markov process on a finite configuration space.

~

We can significantly simplify these expressions for our models, due to the fact that they are one-dimensional, with conservative dynamics in the bulk (no particles are created or destroyed). This means that stationary currents cannot depend on space: there can be no prolonged build-up or depletion of particles anywhere in the system. Perhaps the easiest way to see this formally is through the following procedure.

Consider the diagonal matrix ${\rm e}^{\lambda \tau_i}$ (with $1\leq i\leq L$) with an entry ${\rm e}^{\lambda}$ for all configurations for which site $i$ is occupied, and $1$ otherwise. One may easily check that the matrix similarity ${\rm e}^{-\lambda \tau_i}M_{\mu}{\rm e}^{\lambda \tau_i}$ simply replaces $M_{i-1}(\mu_{i-1})$ and $M_{i}(\mu_{i})$ by, respectively, $M_{i-1}(\mu_{i-1}-\lambda)$ and $M_{i}(\mu_{i}+\lambda)$, and leaves the rest of $M_\mu$ unchanged. That is to say that part of the deformation is transferred from $M_{i-1}(\mu_{i-1})$ to $M_{i}(\mu_{i})$. Using combinations of these transformations for any sites and parameters $\lambda$, we conclude that all the Markov matrices deformed with respect to the currents are similar, and therefore have the same eigenvalues, as long as the sum of the deformation parameters $\sum_{i=0}^{L}\mu_i$ is fixed. Note that the eigenvectors are different, but related to each other through those simple transformations, and that the product of the two eigenvectors, $P_\mu({\mathcal C})\tilde{P}_\mu({\mathcal C})$, is invariant as well.

~

In short, the two quantities that give information on the stationary fluctuations of the current in the system, $E(\mu)$ and $P_\mu({\mathcal C})\tilde{P}_\mu({\mathcal C})$, only depend on the sum of the $\mu_i$'s, which we will write as $\mu$, and all the $j_i$'s have the same value $j$. We may then rewrite eq.(\ref{gEi}) as
\begin{equation}\label{gE}\boxed{
g(j)=j\mu-E(\mu)~~~~{\rm with}~~~~j=\frac{{\rm d}}{{\rm d}\mu} E(\mu)
}\end{equation}
where all the variables are now scalars. The freedom we have in distributing the $\mu_i$'s for a given $\mu$ is quite useful in practice for calculations. In this paper, we will always choose $\mu_i=\mu/(L+1)$.

\subsection{Current and entropy production}
\label{IIc}

A physically meaningful consequence of this simplification, which also highlights the importance of the current as an observable, is that this space-independent current is exactly proportional to the entropy production in the system. Consider the particular set of weights $\{\mu_i\}$ defined by $\mu_i=\nu\log{(p_i/q_i)}$, for which $M_{\mu}$ becomes $M_\nu$:
\begin{equation}\label{MLambda}
m_0=\begin{bmatrix} -p_0 & q_0^{1+\nu}p_0^{-\nu} \\ p_0^{1+\nu}q_0^{-\nu} & -q_0  \end{bmatrix}~,~ M_{i}=\begin{bmatrix} 0 & 0 & 0 & 0 \\ 0 & -q_i & p_i^{1+\nu}q_i^{-\nu} & 0 \\ 0 & q_i^{1+\nu}p_i^{-\nu}& -p_i & 0 \\ 0 & 0 & 0 & 0 \end{bmatrix}~,~m_L=\begin{bmatrix} -q_L & p_L^{1+\nu}q_L^{-\nu} \\  q_L^{1+\nu}p_L^{-\nu} & -p_L  \end{bmatrix}
\end{equation}
which is the deformed Markov matrix measuring the entropy production. We see immediately that
\begin{equation}
M_{-1-\nu}= ~^t\!M_\nu
\end{equation}
which implies the Gallavotti-Cohen symmetry \cite{PhysRevLett.74.2694,Lebowitz99agallavotti-cohen} for the eigenvalues and between the left and right eigenvectors of $M_\nu$ with respect to the transformation $\nu\leftrightarrow(-1\!-\!\nu)$.

Considering that $\mu=\nu \log{\Bigl(\frac{\prod_i p_i}{\prod_i q_i}\Bigr)}$, we also obtain the Gallavotti-Cohen symmetry related to the current, namely
\begin{equation}\boxed{
E(\mu)=E\biggl(-\log{\Bigl(\frac{\prod_i p_i}{\prod_i q_i}\Bigr)}-\mu\biggr)
}
\end{equation}
which is also valid for the other eigenvalues of $M_\mu$, and the corresponding relations between the right and left eigenvectors, as well as a simple relation between the microscopic entropy production $s$, conjugate to $\nu$, and the macroscopic current $j$, conjugate to $\mu$:
\begin{equation}\label{sj}
s= j\log{\Bigl(\frac{\prod_i p_i}{\prod_i q_i}\Bigr)}.
\end{equation}

Two remarks are to be made here. First of all, those weights are ill-defined for the TASEP: micro-reversibility (i.e. the fact that for any allowed transition, the reverse transition is also allowed) is essential to have a fluctuation theorem. Moreover, if we take either the $q_i\rightarrow 0$ or the $L\rightarrow\infty$ limit (with all the $q_i$'s being finitely smaller than the $p_i$'s, so that the logarithm is of order $L$), the centre of the Gallavotti-Cohen symmetry $\mu^\star=-\frac{1}{2}\log{\Bigl(\frac{\prod_i p_i}{\prod_i q_i}\Bigr)}$ is rejected to $-\infty$, so that the `negative current' part of the fluctuations is lost. This will in fact be useful to us: assuming that we can exchange the two limits (the validity of this assumption will be discussed in the conclusion), we will be able to get information about the $j\rightarrow 0$ limit by taking $\mu$ to $-\infty$ in a totally asymmetric situation, which is much simpler than taking it to $-\frac{1}{2}\log{\Bigl(\frac{\prod_i p_i}{\prod_i q_i}\Bigr)}$ in a partially asymmetric one.

Secondly, we may consider the detailed balance case, where $\prod_i p_i=\prod_i q_i$. In that case, there is no entropy production whatsoever, i.e. that $s= 0$, as we see from eq.(\ref{sj}). This does not mean, however, that $j=0$: the deformations through $\mu$ and $\nu$ are in that case not equivalent. The only implication this has on $E(\mu)$ is that it is an even function: $E(\mu)=E(-\mu)$, all the odd cumulants are zero, and positive and negative currents of the same amplitude are equiprobable.

Note that, apart from the precise expressions of $M_\mu$ and $M_\nu$, all of this holds equally well for the case with an extra interaction potential $V$.

\subsection{A note on asymptotics and Legendre transforms}

In the rest of this paper, we will see how analysing the asymptotic behaviour of $M_\mu$ for $\mu\rightarrow\pm\infty$ can give us information on the nature of extreme fluctuations of the current, and indicate the existence of a generic dynamical phase transition. This will be done by obtaining asymptotic expressions or bounds on the generating function of cumulants $E(\mu)$ from diagonalising $M_\mu$ for $\mu\rightarrow\pm\infty$, and taking their Legendre transforms. In order to do that, one has to be careful to check that the Legendre transform of the asymptotic equivalent of $E(\mu)$ is indeed an asymptotic equivalent of the Legendre transform of the real $E(\mu)$. This will be done in appendix \ref{A1}.

~~

Moreover, all calculations will be done at a size $L$ large but finite, so that the limits $\mu\rightarrow\pm\infty$ are taken before $L\rightarrow\infty$. These two limits do not commute in principle, meaning that our results will not be a proof of the existence of a dynamical phase transition but rather a strong indication of it. In particular, in the case of a very inhomogeneous system (possibly with quenched disorder), or a very unphysical potential, the behaviour of $E(\mu)$ and the presence of a sharp dynamical transition will depend on how the $L\rightarrow\infty$ limit is taken on $p_i$ and $V$. On the contrary, for a well-behaved and local $V$, and slow-varying disorder $p_i$, the phase transition is likely to be similar to that which can be observed in the simple ASEP \cite{derrida1999universal,Lazarescu2015}, although it is unclear which of its features are universal. This will be illustrated with a few numerical plots in section \ref{V}.

\newpage

\section{High current limit}
\label{III}

We will first consider the limit where $j\rightarrow\infty$, which corresponds to $\mu\rightarrow\infty$.

As before, we can write
\begin{equation}
M_\mu=M^+_\mu+M^-_\mu+M^d.
\end{equation}
Note that the diagonal part of $M_\mu$ is not deformed. The generic form of the entries of $M^+_\mu$ and $M^-_\mu$ for a given transition and its reverse is
\begin{equation}
{\rm e}^{-V({\cal C})/2}~p_i~{\rm e}^{\mu_i}~{\rm e}^{V({\cal C}')/2}~~~~{\rm and}~~~~{\rm e}^{-V({\cal C}')/2}~q_i~{\rm e}^{-\mu_i}~{\rm e}^{V({\cal C})/2},
\end{equation}
From what we saw in section \ref{IIb}, we can find a function $\phi({\cal C})$ such that the matrices ${\rm e}^{\phi}M^+_\mu{\rm e}^{-\phi}$ and ${\rm e}^{\phi}M^-_\mu{\rm e}^{-\phi}$ have rates
\begin{equation}
{\rm e}^{\frac{\mu}{L+1}}~{\rm e}^{-V({\cal C})/2}~p_i~{\rm e}^{V({\cal C}')/2}~~~~{\rm and}~~~~{\rm e}^{-\frac{\mu}{L+1}}~{\rm e}^{-V({\cal C}')/2}~q_i~{\rm e}^{V({\cal C})/2},
\end{equation}
so that in the $\mu\rightarrow\infty$ limit, $M^d$ and $M^-_\mu$ become negligible, and the problem reduces to the analysis of $M^+_\mu$.

\subsection{Mapping to an XX spin chain}
\label{IIIa}

We are now considering only the right-moving part of $M_\mu$, with entries
\begin{equation}
{\rm e}^{-V({\cal C})/2}~p_i~{\rm e}^{\mu_i}~{\rm e}^{V({\cal C}')/2}
\end{equation}
between configurations which differ by one jump to the right. Using matrix similarities, we can significantly simplify the problem. First of all, it is clear that we can get rid of $V$ through the similarity ${\rm e}^{-V/2}M^+_\mu{\rm e}^{V/2}$, so that $V$ has no influence on large positive fluctuations of the current (as long as $V$ takes finite values). We are left with entries of the form $p_i~{\rm e}^{\mu_i}={\rm e}^{\mu_i+\log(p_i)}$, which is to say that we can treat the inhomogeneity in the rates $p_i$ as an inhomogeneity in the deformations $\mu_i$. Since only their sum matters, only $\prod_i p_i$ will appear in the fluctuations of the current, so that the inhomogeneity is also virtually irrelevant.

~

Putting all this together, we are left with a much simpler matrix to analyse : the right-moving part of a standard ASEP, with all $p_i=1$, and with a pre-factor $({\rm e}^{\mu}\prod_i p_i )^{\frac{1}{L-1}}$. It turns out that it is in fact more convenient to keep a factor $\frac{1}{\sqrt{2}}$ in the boundary matrices, so that we are left to study the matrix
\begin{equation}
M_\mu\sim\Bigl(2{\rm e}^{\mu}\prod_i p_i \Bigr)^{\frac{1}{L-1}}\Biggl(\frac{1}{\sqrt{2}} S_1^+ +\sum\limits_{n=1}^{L-1}S_n^- S_{n+1}^+ +\frac{1}{\sqrt{2}}S_L^-\Biggr)
\end{equation}
where $S_i^+$ and $S_i^-$ are matrices which respectively add and remove a particle at site $i$.

~

We may recognise this to be the upper half of the Hamiltonian of an open XX spin chain \cite{Bilstein1999}. Moreover, it happens to commute with its transpose, thanks to the factors $\frac{1}{\sqrt{2}}$ on each side. We know, from the Perron-Frobenius theorem, that the highest eigenvalue of that matrix is real and non-degenerate. It is therefore also the highest eigenvalue of its transpose, with the same eigenvectors (because they commute). This allows us to define their average $H$, which has the same highest eigenvalue and the same eigenvectors as $M^+$. Forgetting about the constant pre-factor $(2{\rm e}^{\mu}\prod_i p_i )^{\frac{1}{L-1}}$ for the time being, $H$ is given by:
\begin{equation}\boxed{
H=\frac{1}{\sqrt{8}} S_1^x +\frac{1}{2}\sum\limits_{n=1}^{L-1}(S_n^- S_{n+1}^++S_n^+ S_{n+1}^-) +\frac{1}{\sqrt{8}}S_L^x
}\end{equation}
which is the Hamiltonian for the open XX chain with spin $1/2$ and extra boundary terms $S_1^x$ and $S_L^x$ (with $S^x=S^++S^-$). Luckily, we can diagonalise it exactly.

\subsection{Largest eigenvalue of H}
\label{IIIb}

We will not give here the full derivation of the diagonalisation of $H$, which involves standard free fermion techniques, but only the results relevant to this paper. For more details, the reader can refer to section V B of \cite{Lazarescu2015}, or to \cite{Bilstein1999}, where this spin chain is studied with more general boundary conditions.

This system has $2L+2$ independent excitations, with energies 
\begin{equation}
{\cal E}_k=\sin\biggl(\frac{(2k-1)\pi}{2L+2}\biggr)~~~~~~~~{\rm for}~~ k\in\llbracket1,2L+2\rrbracket
\end{equation}
and a vacuum energy of 
\begin{equation}
{\cal E}_0=-\frac{1}{2}\sum\limits_{k=1}^{L+1}{\cal E}_k.
\end{equation}

The highest eigenvalue is therefore obtained by adding all the positive excitation energies, which yields
\begin{equation}
{\cal E}_0+\sum\limits_{k=1}^{L+1}{\cal E}_k=\frac{1}{2}\sin\Bigl(\frac{\pi}{2L+2}\Bigr)^{-1}\sim \frac{L}{\pi}.
\end{equation}
Remembering the global pre-factor $(2{\rm e}^{\mu}\prod_i p_i )^{\frac{1}{L-1}}$ which we removed earlier, and after simplification, we finally get:
\begin{equation}
E(\mu)\sim(\prod_i p_i )^{\frac{1}{L}}~\frac{L}{\pi}~{\rm e}^{\mu/L}
\end{equation}

Depending on how the $p_i$'s are chosen, the first factor in this expression might be negligible or not. However, we can simply decide to always impose that $\prod_i p_i =1$ or is of order $1$, which means choosing the average time it takes for one free particle to go through the system as a natural time scale. Without a great loss of generality, we therefore have
\begin{equation}\boxed{
E(\mu)\sim\frac{L}{\pi}{\rm e}^{\mu/L}
}\end{equation}
and
\begin{equation}\label{IV-3-gMMC}\boxed{\boxed{
g(j)\sim L\bigl( j\log(j)-j(1-\log(\pi))\bigl)
}}\end{equation}
which is proportional to $L$, and {\it depends neither on $V$ nor on the $p_i$'s} (apart from a trivial re-scaling).

This expression, and its scaling with respect to $L$, are what we were mainly after, but it will also be informative to look at the corresponding stationary state.

\subsection{Stationary state conditioned on a high current}
\label{IIIc}

As stated before, the state that we want to examine is the highest energy state of a free fermions system, so that it will not come as a big surprise that the probability of each configuration can be expressed as a Vandermonde determinant. Once more, we will only state the relevant results, and let the curious readers refer to section V B of \cite{Lazarescu2015} for the details of the calculations.

The un-normalised stationary probability of a configuration ${\mathcal C}$ can be expressed, as we saw, as a product of the right and left eigenvectors corresponding to the eigenvalue $E(\mu)$. In the $\mu\rightarrow\infty$ limit, the eigenvectors of $M_\mu$ do not depend on $\mu$ except up to a matrix similarity which leaves this product invariant, so that we will simply write it as ${\rm P}_{\mu\rightarrow\infty}({\mathcal C})$. We have
\begin{equation}
{\rm P}_{\mu\rightarrow\infty}({\mathcal C})={\rm P}\Bigl({\mathcal C}~\Big|~j\!=\!\frac{{\rm d}}{{\rm d}\mu} E(\mu)\Bigr)=P_\mu({\mathcal C})\tilde{P}_\mu({\mathcal C}),
\end{equation}
conditioned on a current $j\sim\frac{1}{\pi}{\rm e}^{\mu/L}$. We find that this probability can be expressed as:
\begin{equation}\boxed{
{\rm P}_{\mu\rightarrow\infty}({\mathcal C})=\prod\limits_{\tau_i=\tau_j}[\sin(r_j - r_i)]\prod\limits_{\tau_i\neq \tau_j}[\sin(r_j + r_i)]
}\end{equation}
where $\tau_i=0$ or $1$ is the occupancy of site $i$, and $r_i=i \pi/(2L+2)$. Note that all these probabilities are still un-normalised.

\begin{figure}[ht]
\begin{center}
 \includegraphics[width=0.5\textwidth]{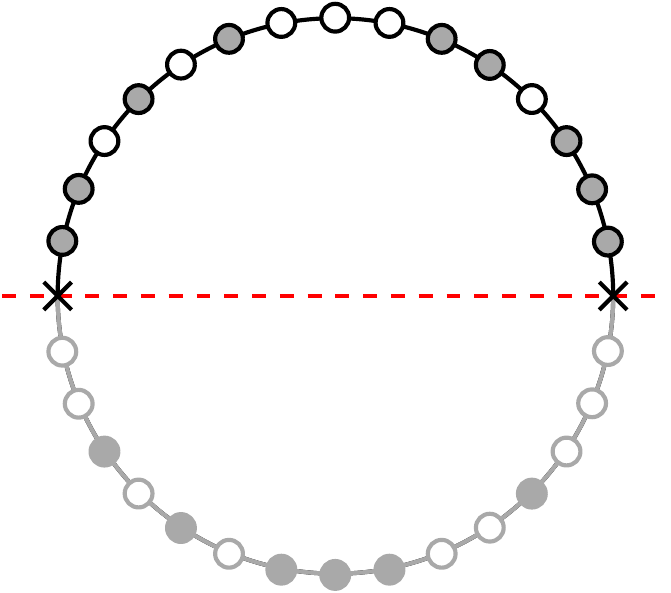}
  \caption{Dyson-Gaudin gas equivalent for the configuration $(110101000110111)$ for the open ASEP conditioned on a large current. The lower part of the system is deduced from the upper part by an axial anti-symmetry.}
\label{fig-DGgas}
 \end{center}
 \end{figure}

This distribution is that of a Dyson-Gaudin gas \cite{Gaudin1973}, which is a discrete version of the Coulomb gas, on a periodic lattice of size $2L+2$, with two defect sites (at $0$ and $L+1$) that have no occupancy, and a reflection anti-symmetry between one side of the system and the other (fig.-\ref{fig-DGgas}). The first (upper) part of the gas is given by the configuration we are considering, and the second (lower) is deduced by anti-symmetry. The interaction potential between two particles at angles $r_i$ and $r_j$ is then given by:
\begin{equation}
V(r_i,r_j)=-\log\bigl(\sin(r_j - r_i)\bigr).
\end{equation}

~

An important feature of that state is the centred density correlation $C_{ij}$ between two sites $i$ and $j$, i.e.
\begin{equation}
C_{ij}=\langle\tau_i\tau_j\rangle-\langle\tau_i\rangle\langle\tau_j\rangle.
\end{equation}
Using the probability distribution we just obtained, we can compute it to be given by
\begin{equation}
C_{ij}=\frac{1}{4(L+1)^2\sin^2\Bigl(\frac{\pi(i+j)}{(2L+2)}\Bigr)}-\frac{1}{4(L+1)^2\sin^2\Bigl(\frac{\pi(i-j)}{(2L+2)}\Bigr)}.
\end{equation}

The correlations are therefore exactly $0$ for sites which are an even number of bonds apart (as is the case for a half-filled periodic chain \cite{Popkov2010}), and behave as
\begin{equation}\boxed{
C_{ij}\sim -\frac{1}{\pi^2 (i-j)^2}
}\end{equation}
otherwise, if the two sites are far away enough from the boundaries. Note that those correlations do not vanish with the size of the system, in contrast with the steady state of the ASEP at $\mu=0$, where they behave as $L^{-1}$ in the maximal current phase and vanish exponentially in the high and low density phases \cite{Derrida1993a}.

~

In the periodic case, it was shown in \cite{Popkov2010} that the large current limit of the steady state of the ASEP of size $L$ converges to a simple periodic Dyson-Gaudin gas (without defects or symmetry). The inhomogeneous interacting version of that model can be treated in the exact same way as we did here, yielding the same result.

We should note that the trick consisting in taking the sum of $M^+_\mu$ and its transpose to reconstruct an XX spin chain is not in fact necessary. All the results can be obtained, in a slightly different way, on $M^+_\mu$ directly, which has the added advantage that the imaginary part of the other eigenvalues is not lost \cite{Karevski2016}.

\newpage

\section{Low current limit}
\label{IV}

We now consider the $j\rightarrow0$ limit.

This case is somewhat more subtle than the previous one. As we saw in section \ref{IIc}, the centre of the Gallavotti-Cohen symmetry is at $\mu^\star=-\frac{1}{2}\log{\Bigl(\frac{\prod_i p_i}{\prod_i q_i}\Bigr)}$, and is also the point where $j=\frac{{\rm d}}{{\rm d}\mu} E(\mu)=0$, by symmetry. Trying to analyse the behaviour of $E(\mu)$ around that point is, in general, not much simpler that the complete problem. However, as we remarked, if we consider the TASEP instead, where all the backward rates are set to $0$, this point becomes $\mu=-\infty$, which greatly simplifies the problem, as we will shortly demonstrate.

The question then remains of the relevance of this special case. Notice that even without taking $q_i=0$, the value of $\mu^\star$ will go to $-\infty$ in the large size limit, as long as the amplitude of the inhomogeneities is bounded (i.e. does not get arbitrarily small or large when $L$ goes to $\infty$), so that $\frac{\prod_i p_i}{\prod_i q_i}$ grows exponentially with $L$. This suggests that the two limits might be consistent (first $L\rightarrow\infty$, then $q_i\rightarrow 0$, or the reverse), but in no way proves it. We will come back on this assumption in the conclusion of this paper, but in all rigour the computations in this section apply only to totally asymmetric models.

~

We will first present the problem, and the tools appropriate for solving it. We will then look at the simple case of the TASEP (which can also be found in \cite{Lazarescu2015} but is reproduced here, in less detail, for pedagogical purposes), where everything can be done fairly explicitly. We will generalise the proof to interacting inhomogeneous systems, where the calculations cannot be done in detail, but where we can still obtain bounds on $E(\mu)$ that will be sufficient for the main result to hold ; that subsection is particularly technical, and the reader interested in the results rather than the methods and proofs may jump directly to section \ref{IVc4}. Finally, we will give a few illustrative examples of cases where the result holds (\ref{IVd1}) or doesn't (\ref{IVd2}). The results and techniques presented in this section constitute the core of this paper, and our main new contribution to the study of dynamical phase transitions.

\subsection{Matrix perturbation and resolvant method}
\label{IVa}

Our starting point here is the deformed Markov matrix for a totally asymmetric model, with all the deformations set to $\varepsilon={\rm e}^{\frac{\mu}{L+1}}$ for simplicity, as in the previous section. In the limit $\mu\rightarrow-\infty$, we have $\varepsilon\rightarrow 0$. We can then write the deformed Markov matrix as
\begin{equation}
M_\mu=M^d+\varepsilon M^+.
\end{equation}

We need to extract the largest eigenvalue of this matrix. Unlike for $\mu\rightarrow\infty$, we cannot simply keep only the leading term, which is this time $M^d$, because it is constant, and therefore not sufficient to give us the behaviour of $g(j)$ through a Legendre transform. We will need to treat the non-diagonal part perturbatively in order to obtain the first correction to that constant term as well.

We will go into the details of how this can be obtained, but the reader familiar with perturbation theory may jump directly to the last paragraph of this sub-section, where these preliminary calculations are summarised.

~

The first step is to find the eigenspace of $M^d$ with the largest eigenvalue. Since $M^d$ is a diagonal matrix containing the opposite of the escape rate from every configuration (i.e. minus the sum of all out-going rates), that eigenspace will be the space of all configurations having the smallest possible escape rate. In other terms, the best states to be stuck into to produce an extremely small current are the ones that have the longest life-time.
Let us write that space as ${\cal S}$, the configurations that live in it as $\{{\cal C}_i^\star\}$, and the corresponding eigenvalue of $M^d$ as $-z_0$. Once we have that space, we need to find the largest eigenvalue of the perturbation $\varepsilon M^+$ restricted to it.

While that perturbative calculation can be performed without too much effort in the simplest cases (where the dominant eigenvalue of $M^d$ is not degenerate), we will need to cover all possible complications. This can be done in a very compact way using the so-called \textit{resolvent formalism} \cite{Fredholm1903}. Not only is it a clean and systematic way to deal with perturbative expansions in degenerate spaces, but it also provides with a rigorous definition of an effective dynamics within that space.

This formalism can be stated as follows: for a diagonalisable matrix $A$ with eigenvalues $E_i$ and eigenvectors $ |P_i\rangle$ and $\langle P_i|$, we may write
\begin{equation}\label{IV-3-Eproj}
\oint_{C}\frac{dz}{{\rm i} 2\pi}\frac{z}{z-A}=\sum\limits_{E_i\in C} E_i|P_i\rangle\langle P_i|
\end{equation}
where the sum is over the eigenvalues of $A$ which lie inside of the contour $C$.

Since we are only interested in the eigenvalues of $M_\mu$ which are close to the dominant eigenvalue $-z_0$ of $M^d$, we can apply this formula to $M_\mu$ with a small contour around $-z_0$ to get our effective biased Markov matrix, which we will write as $-z_0+M_{eff}$ (keeping the scalar term out for convenience). Shifting the origin of the complex plane to $-z_0$ for simplicity, we get
\begin{equation}\label{IV-3-Eproj2}
M_{eff}=\oint_{C}\frac{dz}{{\rm i} 2\pi}\frac{z}{z-z_0-M^d-\varepsilon M^+}
\end{equation}
where $C$ is a small circle centred at $0$. We can now expand this expression in powers of $\varepsilon$:
\begin{equation}\label{MeffPath}
M_{eff}=\oint_{C}\frac{dz}{{\rm i} 2\pi} \sum\limits_{k=0}^{\infty}\frac{z}{z-(M^d+z_0)}\Bigl(M^+ \frac{1}{z-(M^d+z_0)} \Bigr)^k \varepsilon ^k
\end{equation}
which is a sum over paths of length $k$, with transitions given by $M^+$ and a configuration weight given by $(z-M^d-z_0)^{-1}$. We see that, if $C$ is small enough to contain only the poles of $(z-M^d-z_0)^{-1}$ which are at $0$, the only terms which contribute to the integral (i.e. that give first order poles which yield non-zero residues) are those for which $M^d$ is taken at $-z_0$ at least twice. In other terms, they are the paths which go through at least two of the ${\cal C}_i^\star$'s.

Moreover, since we are only interested in the leading term in the largest eigenvalue of $M_{eff}$, the corrections to the eigenvectors will not be relevant, so that $M_{eff}$ is equivalent to its projection onto ${\cal S}$. This means that we can restrict the expression (\ref{MeffPath}) to paths that start and end in ${\cal S}$, from some ${\cal C}_i^\star$ to ${\cal C}_j^\star$, accounting for the previous requirement. 

Finally, notice that the weight of each of those paths is proportional to $\varepsilon$ to the power of the number of jumps performed, with a pre-factor which is the ratio of all the jump rates used on that path over all the escape rates of the intermediate configurations minus $z_0$. Since we are only interested in the leading order in $\varepsilon$, we only need to consider the paths with the least number of jumps between the initial and final configuration. The corresponding entry in $M_{eff}$ will have a pre-factor, being the sum of the ones for each of the paths that have that minimal length, and which we can calculate in the simplest cases, but which will ultimately be inessential to our result.

~

Let us summarise this first section before getting into specific calculations. If $M^d$ has a highest eigenvalue $-z_0$ with eigenspace ${\cal S}$ generated by configurations $\{{\cal C}_i^\star\}$, then the highest eigenvalue of $M_\mu$ is given by $-z_0$ plus a first correction which is the largest eigenvalue of a matrix $M_{eff}$ with entries $A_{ij}\varepsilon^{d_{ij}}$, where $d_{ij}$ is the smallest number of jumps connecting ${\cal C}_j^\star$ to ${\cal C}_i^\star$, and $A_{ij}$ is a numerical pre-factor which can be obtained from expression (\ref{MeffPath}) although it will not be necessary.

\subsection{TASEP conditioned on low current}
\label{IVb}

We first look at the simple case of the regular TASEP, with all bulk rates equal to $p$, and boundary rates $p_0$ and $p_L$. It is customary to consider $p_0\leq p$ and $p_L\leq p$, which does not restrict the behaviour of the system by much. We will focus on those case and merely comment on the remaining ones, which are covered in all generality by section \ref{IVc}.

~

For this model, we can easily narrow down the set of candidates for ${\cal S}$: since all the transition rates are independent (i.e. the rate of a jump does not depend on which other jumps are possible), the escape rate of a state with several possible jumps is the sum of escape rates of states which have only one possible jump, and is therefore larger than each one of those. We only need to consider states with a single allowed jump, which is to say states that are completely full up to a given site and then completely empty.

\begin{figure}[ht]
\begin{center}
 \includegraphics[width=0.4\textwidth]{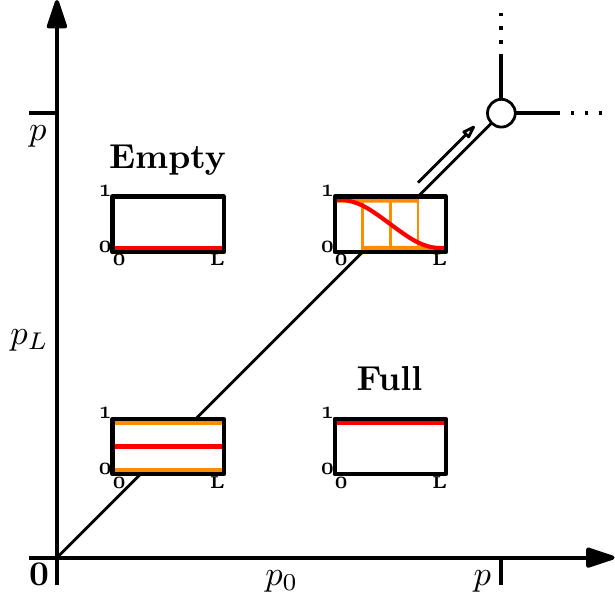}
  \caption{Phase diagram of the open ASEP for very low current. The mean density profiles are represented in red the insets. The profiles in orange are the individual configurations which compose the steady state.}
\label{fig-DiagLC}
 \end{center}
 \end{figure}

We then see that there are three qualitatively different situations (fig.-\ref{fig-DiagLC}):
\begin{itemize}
\item if $p_0<p_L$ and $p_0<p$, then the best configuration is empty ($\tau_i=0$ for all $i$'s), with $z_0=p_0$. If $p_L<p_0$ and $p_L<p$, we have the same in reverse: the best configuration is full ($\tau_i=1$ for all $i$'s) and $z_0=p_L$ (those two first cases are symmetric to one another, and we will only be considering the first one) ;
\item if $p_0=p_L<p$, then $z_0=p_0$, and we have two competing configurations: empty or full ;
\item if $p_0\geq p$ and $p_L\geq p$, any configuration with a block of $1$'s followed by a block of $0$'s has an eigenvalue of $-z_0=-p$, which is the highest, except possibly the completely full and empty configurations depending on whether those inequalities are strict or saturated ; all those situations being essentially identical in the large size limit, we will focus on $p_0=p_L=p$, in which case there are $L+1$ states in ${\cal S}$.
\end{itemize}

We will now construct $M_{eff}$ in each of these cases, and analyse its largest eigenvalue.

\subsubsection{Empty/full phases}
\label{IV-1-a}

This first case, where $z_0=p_0 <p_L$, is fairly straightforward: ${\cal S}$ contains only the empty configuration, and $M_{eff}$ has only one entry, equal to the weight of a path going from the empty configuration to itself. That is achieved in the least number of steps, which is $L+1$, by having one single particle go through the whole system from left to right. The contribution of $\varepsilon$ is therefore $\varepsilon^{L+1}={\rm e}^{\mu}$, and the numerical pre-factor is simply given by $\frac{p_0}{p-p_0}$.

This gives us
\begin{equation}\label{IV-3-EmuLD}
E(\mu)\sim -p_0+{\rm e}^{\mu}\frac{p_0}{p-p_0}
\end{equation}
and
\begin{equation}\label{IV-3-gjLD}\boxed{
g(j)=p_0+j\log(j)-j\bigl(\log(p_0/(p-p_0))+1\bigr).
}\end{equation}

Finally, note that in this case, the second largest eigenvalue is $-p_L$ for $\varepsilon\rightarrow 0$ (and corresponds to a completely full system), so that the gap between the first two eigenvalues of $M_\mu$ is finite and equal, at leading order, to $\Delta E=(p_L-p_0)$. 

The corresponding results for $p_L<p_0$ (the `full' phase) can be obtained by exchanging $p_0$ with $p_L$.

\subsubsection{Empty-full line}
\label{IV-1-b}

We now consider the slightly more complex case where $p_0=p_L<p$. This time, there are two states with equal eigenvalues for $\mu=-\infty$:
\begin{equation}\label{IV-3-P0}
|0\rangle=|00\dots00\rangle~~~~{\rm and}~~~~|1\rangle=|11\dots11\rangle.
\end{equation}

As in the previous case, the diagonal part of $M_{eff}$ is given by the shortest paths going from these configurations to themselves, with the same weight as before (them being equal to each other because of the particle-hole symmetry). For the off-diagonal elements, we have to consider the shortest way to go from $|0\rangle$ to $|1\rangle$, or the opposite. This means completely filling or emptying the system, which can be done in $L(L+1)/2$ steps, but in this case the pre-factor contains contributions from many paths and is not straightforward to calculate. We therefore have something of the form
\begin{equation}
M_{eff}=\begin{bmatrix} {\rm e}^{\mu}\frac{p_0}{p-p_0} &X {\rm e}^{\frac{L}{2}\mu}\\ X {\rm e}^{\frac{L}{2}\mu} &{\rm e}^{\mu}\frac{p_0}{p-p_0}  \end{bmatrix}
\end{equation}
where $X$ is said pre-factor.

From this, we see that the difference between the two highest eigenvalues is of order $\varepsilon^{L(L+1)/2}={\rm e}^{\frac{L}{2}\mu}$. For symmetry reasons, the main eigenvector is then $\frac{1}{2}(|0\rangle+|1\rangle)$, and the second one is $\frac{1}{2}(|0\rangle-|1\rangle)$. The leading terms of the largest eigenvalue are the same as before:
\begin{equation}\label{IV-3-EmuLDHD}
E(\mu)\sim -p_0+{\rm e}^{\mu}\frac{p_0}{p-p_0}
\end{equation}
and
\begin{equation}\label{IV-3-gjLDHD}\boxed{
g(j)=p_0+j\log(j)-j\bigl(\log(p_0/(p-p_0))+1\bigr)
}\end{equation}
but this time, the gap behaves as $\Delta E\sim{\rm e}^{\frac{L}{2}\mu}$.

\subsubsection{Anti-shock zone}
\label{IV-1-c}

For the last case, where all the jumping rates are equal ($p_0=p_L=p$), we find $L+1$ states with an eigenvalue equal to $-p$ for $\mu=-\infty$. Those states are given by $|k\rangle=|\{1\}_{(k)} \{0\}_{(L-k)}\rangle$, i.e. configurations made of a block of $1$'s followed by a block of $0$'s. Those are called \textit{anti-shocks}, being symmetric to the usual shocks observed in the steady state of the TASEP for $p_0=p_L\leq p/2$, which have a low density region followed by a high density one.

Using the resolvent formalism, we find:
\begin{align}
\langle k| M_{eff} |k\rangle&\sim\varepsilon^{L+1},\\
\langle k+1| M_{eff} |k\rangle&\sim\varepsilon^{k+1},\\
\langle k-1| M_{eff} |k\rangle&\sim\varepsilon^{L-k+1},
\end{align}
as well as terms of the type
\begin{align}
\langle k+2| M_{eff} |k\rangle&\sim X\varepsilon^{2k+3},\\
\langle k-2| M_{eff} |k\rangle&\sim Y\varepsilon^{2L-2k+3},\\
\langle k+3| M_{eff} |k\rangle&\sim Z\varepsilon^{3k+6},
\end{align}
and so on. We can check those last terms to be of sub-leading order in $E(\mu)$, and we will neglect them right away, which will allow us to complete our calculation. A more rigorous approach will be taken in section \ref{IVc}.

We are left with
\begin{equation}
M_{eff}=p\varepsilon^{L+1}+p\sum\limits_{k=1}^{L}\Bigl(\varepsilon^{k}|k\rangle\langle k-1|+\varepsilon^{L-k+1}|k-1\rangle\langle k|\Bigr).
\end{equation}
This matrix is similar to
\begin{equation}
M_{eff}=p\varepsilon^{L+1}+p\varepsilon^{(L+1)/2}\sum\limits_{k=1}^{L}\Bigl(|k\rangle\langle k-1|+|k-1\rangle\langle k|\Bigr)
\end{equation}
and can be diagonalised exactly (see \cite{Lazarescu2015} for more details). This time the dominant contribution to the largest eigenvalue turns out to come from the non-diagonal part, and be equal to $-2p\varepsilon^{(L+1)/2}\cos(\pi/(L+2))$, which yields
\begin{equation}\label{IV-3-EER-}
E(\mu)\sim-p+2p~{\rm e}^{\mu/2}
\end{equation}
and
\begin{equation}\label{IV-3-gjLDc}\boxed{
g(j)=p+2j\log(j)-2j.
}\end{equation}
In this case, the gap is equal to
\begin{equation}
\Delta E=2p\varepsilon^{(L+1)/2}\Bigl(\cos\bigl(\pi/(L+2)\bigr)-\cos\bigl(2 \pi/(L+2)\bigr)\Bigr)\sim  \frac{3p\pi^2}{L^2}{\rm e}^{\mu/2}.
\end{equation}

The cases where $p_0>p$ and/or $p_L>p$ are almost identical, with the first state $|0\rangle$ and/or the last state $|L\rangle$ removed, and show the same large-size behaviour.

\subsection{Interacting inhomogeneous TASEP}
\label{IVc}

In this section, which contains the main new result of the present paper, we generalise the ones obtained for the standard TASEP to the inhomogeneous and interacting processes defined in section \ref{II}. That result can be stated as follows: in the limit of small currents ($\mu\rightarrow-\infty$), the first correction to the generating function of the cumulants of the current scales exponentially with $\mu$, with a rate that does not vanish in the limit of large sizes $L\rightarrow\infty$ (as it does in the high current limit):
\begin{equation}
\exists \{A,B\}>0: ~~~~ E(\mu)+z_0\sim{\rm e}^{B\mu}\ll {\rm e}^{A\mu}~~~~{\rm for}~~~~\mu\rightarrow -\infty,~~L\rightarrow\infty
\end{equation}
where $B$ can depend on $L$ but not $A$. 

The fact that it scales exponentially with $\mu$ simply comes from it being an eigenvalue of a finite matrix $M_{eff}$ whose elements are powers of $\varepsilon$ (with unimportant numerical pre-factors). The proof of the bound can then be achieved in three steps:
\begin{itemize}
\item we will first show that, under certain assumptions, the states in $\mathcal{S}$, which have the longest lifetime, are similar to anti-shocks, with a region transiting from completely full to completely empty which can have any occupancy but whose size cannot grow with $L$ ; 
\item we will then show that a simple cycle of such states takes a number of steps that is at least one order of $L$ larger than the number of states in the cycle ;
\item finally, we will use that bound to estimate the principal minors of $M_{eff}$, in order to show that the leading power of $\varepsilon$ in the largest eigenvalue of $M_{eff}$ is linear in $L$, which will complete the proof.
\end{itemize}
Note that all of the estimates we will be making are broad enough to account for the worst cases, but can certainly be made more precise for specific models.

\subsubsection{Longest-lived states}
\label{IVc1}

We will first show that all the states in $\mathcal{S}$ are of the form $|\{1\}_{(k^-)} \{\tau_i\}_{(k^+-k^-)}\{0\}_{(L-k^+)}\rangle$ with $k^+-k^-\leq K$ where $K$ is independent of $L$, i.e. anti-shocks with a maximal width $K$. In this expression, $k^-$ is the site of the first particle that can jump, and $k^+$ that of the last one (it is assumed that site $k^-+1$ is empty and that site $k^+$ is occupied, unless $k^-=k^+$).

In order to prove that first step, we need to make two assumptions:
\begin{itemize}
\item first, that the values of the inhomogeneous rates do not scale with the size of the system, i.e. that the values of $p_i$ are bounded on both sides for any value of $L$:
\begin{equation}\label{pbound}
\exists \{p_{min},p_{max}\},\forall L,\forall i: ~~~~0< p_{min}<p_i<p_{max}
\end{equation}
\item secondly, that the potential $V$ is such that any difference of $V$ over a transition (i.e. $V(\mathcal{C}')-V(\mathcal{C})$ if $\mathcal{C}\sim \mathcal{C}'$) is bounded on both sides:
\begin{equation}\label{vbound}
\exists \{\delta V_{min},\delta V_{max}\},\forall L,\forall \mathcal{C}\sim \mathcal{C}':~~~~2\delta V_{min}< V(\mathcal{C}')-V(\mathcal{C})< 2\delta V_{max},
\end{equation}
and that it is local, in the sense that the potential difference for a jump between sites $i$ and $i+1$ doesn't depend strongly on the part of the configuration that is far enough from the jump:
\begin{align}\label{vlocal}
\forall \alpha> 0, &~\exists~l_{\alpha}, \forall L, \forall i,\forall \mathcal{C}_1\sim \mathcal{C}_1',\forall \mathcal{C}_2\sim \mathcal{C}_2':\\
&\mathcal{C}_1|_{[i-l_\alpha,i+1+l_\alpha]}=\mathcal{C}_2|_{[i-l_\alpha,i+1+l_\alpha]}~{\rm and}~\mathcal{C}'_1|_{[i-l_\alpha,i+1+l_\alpha]}=\mathcal{C}'_2|_{[i-l_\alpha,i+1+l_\alpha]}\nonumber\\
&~~~~~~\Rightarrow~~|V(\mathcal{C}_1')-V(\mathcal{C}_1)-V(\mathcal{C}_2')+V(\mathcal{C}_2)|<2\log(1+\alpha)\nonumber
\end{align}
where $\mathcal{C}|_{[i-l_\alpha,i+1+l_\alpha]}$ is the portion of $\mathcal{C}$ that is between sites $i-l_\alpha$ and $i+1+l_\alpha$, and the $2\log$ is there for later convenience ; these two conditions are for instance verified if $V$ is a short-range two-body potential, and are not verified for typical long-range potentials, but are less restrictive than that.
\end{itemize}

From assumptions (\ref{pbound}) and (\ref{vbound}), we get that the transition rates $w(\mathcal{C},\mathcal{C}')$ are bounded independently of $L$:
\begin{equation}
w_{min}=p_{min}~{\rm e}^{\delta V_{min}}<w(\mathcal{C},\mathcal{C}')<p_{max}~{\rm e}^{\delta V_{max}}=w_{max}.
\end{equation}

~~

We can then prove the following crucial statement: any state with the minimal escape rate $z_0$ has at most $\bigl\lfloor w_{max}/w_{min}\bigr\rfloor$ possible transitions. The proof is straightforward: if a longest-lived state $\mathcal{C}$ has $n_t$ possible transitions, then, from the left side of the previous inequality, its escape rate is larger than $n_t~w_{min}$ ; from the right side of the inequality, it also has to be smaller than $w_{max}$, because there are, for instance, states with only one transition whose escape rate is one single jump rate and which is smaller than that bound ; therefore
\begin{equation}
n_t< \frac{w_{max}}{w_{min}}.
\end{equation}

~~

\begin{figure}[ht]
\begin{center}
 \includegraphics[width=0.5\textwidth]{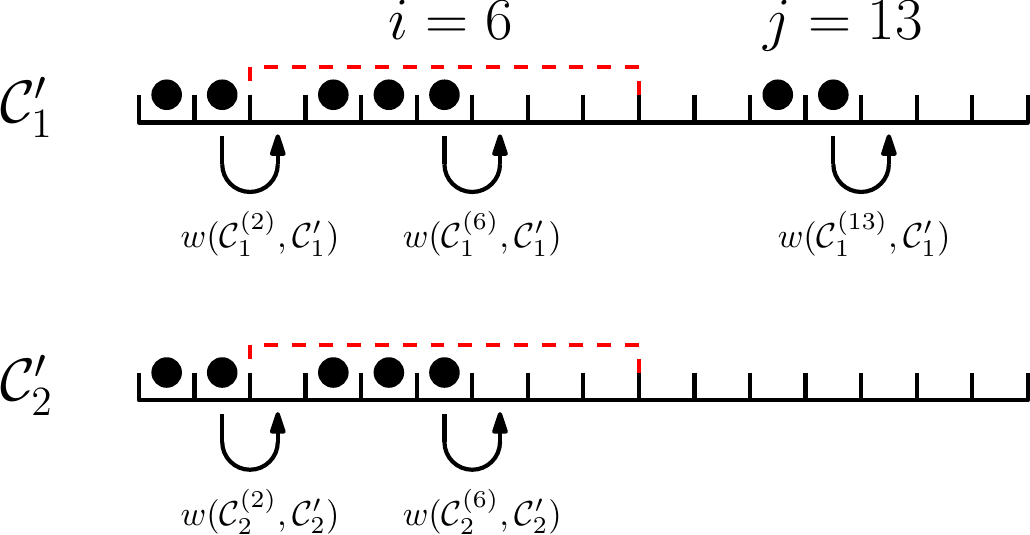}
  \caption{Example of configurations $\mathcal{C}'_1$ and $\mathcal{C}'_2$ with $l=3$, $i=6$ and $j=13$. The area in the dashed red box represents the part of those configurations which could have a strong influence on the value of $w(\mathcal{C}_1^{(6)},\mathcal{C}'_1)$ or $w(\mathcal{C}_2^{(6)},\mathcal{C}'_2)$, and which is identical in both configurations.}
\label{fig-Distance}
 \end{center}
 \end{figure}

Moreover, those $n_t$ transitions cannot be too far from each other, because of the locality assumption (\ref{vlocal}) which ensures that particles cannot stabilise each-other from afar. Consider for instance a configuration $\mathcal{C}'_1$ with two successive possible jumps involving particles at sites $i$ and $j$ such that $j-i>2l$. This means that the first of the two particles has more than $l$ holes in front of it, or that the second has more than $l$ particles behind it. Let us assume it is the former (but the latter can be treated in the exact same way). Consider also the configurations $\mathcal{C}_1^{(k)}$ resulting from one of the jumps, from site $k$ to $k+1$, the configuration $\mathcal{C}'_2$ which is the same as $\mathcal{C}'_1$ up to site $i$ and empty  afterwards, and the configurations $\mathcal{C}_2^{(k)}$ resulting from the jump from site $k$ to $k+1$ in $\mathcal{C}'_2$ (cf. fig.\ref{fig-Distance}). $\mathcal{C}'_2$ has at most $n_t-1$ possible jumps. 

Choosing now $l=l_\alpha$ for $\alpha=(w_{min}/w_{max})^2$, we have, according to (\ref{vlocal}):
\begin{equation}
w(\mathcal{C}_2^{(k)},\mathcal{C}'_2)<(1+\alpha)w(\mathcal{C}_1^{(k)},\mathcal{C}'_1)~~~~{\rm for}~~~~k\leq i
\end{equation}
because $\mathcal{C}'_1$ and $\mathcal{C}'_2$ are identical up to a distance $l_\alpha$ from any transitions that they have in common. We can now compare the escape rates from $\mathcal{C}'_1$ and $\mathcal{C}'_2$:
\begin{equation}
\sum\limits_k w(\mathcal{C}_2^{(k)},\mathcal{C}'_2)<(1+\alpha)\sum\limits_{k\leq i} w(\mathcal{C}_1^{(k)},\mathcal{C}'_1)<\sum\limits_{k\leq i} w(\mathcal{C}_1^{(k)},\mathcal{C}'_1)+(n_t-1)~\alpha~ w_{max}<\sum\limits_{k\leq i} w(\mathcal{C}_1^{(k)},\mathcal{C}'_1)+w_{min}.
\end{equation}
The right-hand side is smaller than the escape rate from $\mathcal{C}'_1$, since there is at least one transition to the right of $i$. Therefore, it cannot be one of the longest-lived states, because $\mathcal{C}'_2$ has a strictly longer lifetime.

From this, we conclude that longest-lived states cannot have transitions that are further apart than $2 l_\alpha$ with $\alpha=(w_{min}/w_{max})^2$.

~~

Combining this with the previous result, we find that all the transitions from a longest-lived state have to be within a region of size $K=2 l_\alpha \bigl(\bigl\lfloor w_{max}/w_{min}\bigr\rfloor-1\bigr)$, which is independent of $L$. All the sites to the left of that region have to be full, and all those to the right have to be empty. This concludes this first part of the proof.

\subsubsection{Length of simple cycles in $\mathcal{S}$}
\label{IVc2}

For this second step, we will be estimating the minimal number of jumps performed along a simple cycle of $n$ states from $\mathcal{S}$: $\mathcal{C}_1\rightarrow\mathcal{C}_2\rightarrow\dots\rightarrow\mathcal{C}_n\rightarrow\mathcal{C}_1$, where no state is visited more than once. Let us write $d=\sum\limits_{i\rightarrow j} d_{ji}$ the total minimal number of steps along that cycle. As we saw in the previous section, the states in $\mathcal{S}$ can be indexed by the position $k^-$ of the first jump, the position $k^+$ of the last jump (with $k^+-k^-\leq K$), and the configuration $\{\tau_i\}$ of the sites in between. We will also write the number of particles in a state as $N_{\mathcal{C}}=k^-+\sum\tau_i$.

We want to show that there is a constant $A$ independent of $L$ such that $d>(L+1)~n~A$.

~~

Consider first a cycle of any length, however small. Since we are coming back to the initial state at the end, the total number of jumps has to be a multiple of $L+1$ (the jumps can be reordered so that every particle does an integral number of loops around the system before coming back to its initial position). Since at least one step is taken, we have then $d>L$. This is true in particular for cycles of length $1$, so that $d_{ii}>L$ for any $i$.

To obtain a bound on the number of steps of a larger cycle, we can first simplify the problem by noticing that any state $\mathcal{C}$ is a finite number of steps away from the totally ordered state with as many particles, which we will call $\mathcal{C}^{(o)}$, with $N_{\mathcal{C}}$ particless followed by $L-N_{\mathcal{C}}$ holes (i.e. the state that we called $|N_{\mathcal{C}}\rangle$ for the simple TASEP). In the worst case, it takes $K^2/4$ steps to go from $\mathcal{C}^{(o)}$ to $\mathcal{C}$, which happens if $K$ is even and if $\mathcal{C}$ is of the form $|\{1\}_{(k^-)} \{0\}_{(K/2)}\{1\}_{(K/2)}\{0\}_{(L-k^--K)}\rangle$. Considering now the number of steps from $\mathcal{C}_1^{(o)}$ to $\mathcal{C}_2^{(o)}$, we know that it is larger than that from $\mathcal{C}_1^{(o)}$ to $\mathcal{C}_1$ plus that from $\mathcal{C}_1$ to $\mathcal{C}_2$ plus that from $\mathcal{C}_2$ to $\mathcal{C}_2^{(o)}$. Writing $d_{ij}^{(o)}$ as the minimal number of steps from $\mathcal{C}_j^{(o)}$ to $\mathcal{C}_i^{(o)}$, we have therefore that
\begin{equation}
d_{ij}>d_{ij}^{(o)}-\frac{K^2}{2}.
\end{equation}

Moreover, depending on $N_{\mathcal{C}_1}$ and $N_{\mathcal{C}_2}$, it is straightforward to obtain $d_{21}^{(o)}$: 
\begin{itemize}
\item if $N_{\mathcal{C}_1}>N_{\mathcal{C}_2}$, then the last $N_{\mathcal{C}_1}-N_{\mathcal{C}_2}$ particles have to leave the system through the right boundary, which can be done with
\begin{equation}
d_{21}^{(o)}=\frac{1}{2}(2L+1-N_{\mathcal{C}_1}-N_{\mathcal{C}_2})(N_{\mathcal{C}_1}-N_{\mathcal{C}_2}).
\end{equation}
\item if $N_{\mathcal{C}_1}<N_{\mathcal{C}_2}$, then the last $N_{\mathcal{C}_2}-N_{\mathcal{C}_1}$ particles have to enter the system from the left boundary, which can be done with
\begin{equation}
d_{21}^{(o)}=\frac{1}{2}(1+N_{\mathcal{C}_1}+N_{\mathcal{C}_2})(N_{\mathcal{C}_2}-N_{\mathcal{C}_1}).
\end{equation}
\end{itemize}
Note that in all cases, $d_{21}^{(o)}+d_{12}^{(o)}=(L+1)|N_{\mathcal{C}_2}-N_{\mathcal{C}_1}|$, so that
\begin{equation}
d_{21}+d_{12}>(L+1)|N_{\mathcal{C}_2}-N_{\mathcal{C}_1}|-K^2.
\end{equation}

It follows that, for a cycle with $N^-=\min[\{N_{\mathcal{C}_i}\}]$ and $N^+=\max[\{N_{\mathcal{C}_i}\}]$, the total number of steps has to be larger than the direct path back and forth between two states that realise those extrema, so that
\begin{equation}\label{dNN}\boxed{
d>(L+1)(N^+-N^-)-K^2.
}\end{equation}

One final remark to be made is that a certain value of $N_{\mathcal{C}}$ can correspond to at most $2^{K-1}$ different states from $\mathcal{S}$. This can be seen by considering a state with $k^-=N_{\mathcal{C}}-n$, with the other $n$ particles being confined to the $K-1$ sites following the first hole at $k^-+1$. There are at most ${K-1 \choose n}$ such states, for $n$ from $1$ to $K-1$, plus the state with $n=0$, which all sum up to $2^{K-1}$.

~~

Consider now a cycle of length $n>2^{K+1}$. From what we just saw, $N_{\mathcal{C}}$ takes at least $n~2^{-K+1}$ different values along the cycle, so that $N^+-N^->n~2^{-K+1}-1=4n~2^{-K-1}-1>n~2^{-K-1}+2$. Combining this with the previous inequality (\ref{dNN}) gives
\begin{equation}
d>(L+1)(n~2^{-K-1}+2)-K^2,
\end{equation}
which, for $L+1>K^2/2$, finally gives
\begin{equation}\boxed{
d>(L+1) ~n~A
}\end{equation}
with $A=2^{-K-1}$.

\subsubsection{Equivalent and bound for $E(\mu)$}
\label{IVc3}

We will now find a bound on the eigenvalues of $M_{eff}$ using the bounds we have on the weight of its cycles. This can be done by looking at the characteristic polynomial of $M_{eff}$:
\begin{equation}
P_\varepsilon(x)={\rm det}\bigl[x~\delta_{ij} -A_{ij}\varepsilon^{d_{ij}} \bigr]=\sum_{k=0}^{N} a_k~x^{N-k}
\end{equation}
where $N=|\mathcal{S}|$ and $a_0=1$. It is well known that $a_k$ can be expressed as a sum of the principal minors of size $k$ of $M_{eff}$, which is to say a sum of weights of all composite cycles from $M_{eff}$ of total length $k$. Each $a_k$ is therefore a polynomial in $\varepsilon$, the valuation (smallest exponent) of which we will write as $m_k$. Let us also define
\begin{equation}\label{mindn}
C= \min_{k:1..N}\biggl[\frac{m_{k}}{k}\biggr]=\min_{\rm cycles}\biggl[\frac{d}{n}\biggr],
\end{equation}
where, as in the previous section, $d$ is the number of steps in a cycle of length $n$. We have shown that $C>(L+1)A$.

Consider now the rescaled polynomial
\begin{equation}
Q_\varepsilon(x)=\varepsilon^{-N C} P_\varepsilon(\varepsilon^{C}x)=\sum_{k=0}^{N} a_k\varepsilon^{-kC}~x^{N-k}
\end{equation}
which has at least one finite coefficient $a_k\varepsilon^{-kC}$ ($k\neq 0$), all the others being infinitesimal in $\varepsilon$. The roots of this polynomial are therefore finite in the limit $\varepsilon\rightarrow 0$, and at least one of them is non-vanishing. It is not obvious that the highest root, which is the one we are interested in, is among those roots, as they could in principle be all negative. We will however see that it is the case here.

Consider the matrix $\tilde{M}_{eff}$ where only the entries that contribute to the cycles that realise the minimum (\ref{mindn}) are kept:
\begin{equation}
\tilde{M}_{eff}=\sum\limits_{i\sim j}A_{ij}\varepsilon^{d_{ij}},
\end{equation}
where $i\sim j$ means that the transition $j\rightarrow i$ is on at least one cycle such that $d=nC$. In particular, for transitions which can be done in the same number of steps through different paths, only the paths with the most intermediate states (the highest $k$) will be kept, which explains why some terms were sub-dominant in $M_{eff}$ in section \ref{IV-1-c}. By construction, the characteristic polynomial of $\tilde{M}_{eff}$ is $\varepsilon^{N C}Q_0(\varepsilon^{-C}x)$: we can obtain it by rescaling $P_\varepsilon$, taking $\varepsilon$ to $0$ so that only the finite terms survive, and finally taking it back to the original scaling. Consider also $\hat{M}_{eff}=\tilde{M}_{eff}\big|_{\varepsilon=1}$
\begin{equation}
\hat{M}_{eff}=\sum\limits_{i\sim j}A_{ij},
\end{equation}
whose characteristic polynomial is therefore $Q_0(x)$. Both $\hat{M}_{eff}$ and $\varepsilon^{-C}\tilde{M}_{eff}$ are diagonalisable, because all of their entries are on at least one cycle, and they have the same characteristic polynomial, from which we conclude that they are similar. In particular, the largest eigenvalue of $\varepsilon^{-C}\tilde{M}_{eff}$ is the same as that of $\hat{M}_{eff}$, which is a positive matrix. It is therefore strictly positive.

Moreover, the matrix $\hat{M}_{eff}$ contains all the information relative to the effective process at low current. We will describe that process in two simple cases in section \ref{IVd1} (fig.\ref{fig-Effective}).

~~

Going back to $M_{eff}$, we conclude that its largest eigenvalue scales as $\varepsilon^{C}\sim {\rm e}^{B\mu}$ with
\begin{equation}\boxed{
B=\frac{C}{L+1}\ll A
}\end{equation}
for $\mu\rightarrow-\infty$. This concludes our proof.

\subsubsection{Conclusion: asymptotics of the large deviation function of the currents}
\label{IVc4}

We conclude by re-stating the assumptions and the result from this section, and examining its consequence on the large deviation function of the current.

We have seen that, for a generalised TASEP with bounded inhomogeneities and a short-range potential, we can find a quantity $A$ independent of the size of the system such that 
\begin{equation}
|E(\mu)+z_0|\ll {\rm e}^{A\mu}
\end{equation}
for $L\rightarrow\infty$ and $\mu\rightarrow-\infty$. Moreover, $E(\mu)+z_0$ is equivalent to the largest eigenvalue of a matrix $M_{eff}$ who behaves as
\begin{equation}\boxed{
E(\mu)+z_0\sim {\rm e}^{B\mu}
}\end{equation}
for $\mu\rightarrow-\infty$ with $B$ bounded as a function of $L$. The states in $\mathcal{S}$, which are the ones involved in the dynamics in that limit, are anti-shocks, of the form $|\{1\}_{(k^-)} \{\tau_i\}_{(k^+-k^-)}\{0\}_{(L-k^+)}\rangle$ with $k^+-k^-\leq K$ where the maximal width $K$ of the anti-shock is independent of $L$

The statements and proofs did not require any specific form for the inhomogeneous rates and the potential, other than conditions (\ref{pbound}-\ref{vlocal}), which means that there is no reason for $B$ to have a limit for $L\rightarrow\infty$. However, if $V$ is local and well-behaved and $p_i$ is slowly varying in space, we can expect that limit to exist.

Having then an equivalent of that form, we can deduce that
\begin{equation}\label{LowCurrtEquiv}\boxed{\boxed{
g(j)\sim z_0+B^{-1}j\log(j)
}}\end{equation}
which does not scale with the size of the system. This and the localised nature of the states in $\mathcal{S}$, which are the typical states occupied by the system at low current, are compatible with a hydrodynamic description of the fluctuating current \cite{Bodineau2006,Lazarescu2015}, where the cost of maintaining such a fluctuation in the system comes from one localised defect and hence is independent of $L$.

\subsection{Illustrative examples}
\label{IVd}

In this section, we give a few simple examples to illustrate our result beyond the excessively simple case of the TASEP. We start with a family of models with finite-range interaction for which the width $K$ of the anti-shock region can be tuned to any value, and the number of relevant states can be adjusted as well. We also explicitly build the effective dynamics $M_{eff}$ in the simplest non-trivial case. Finally, to illustrate the necessity of having local interactions, we give an example of a system with long-range interactions where our result does not hold.

\subsubsection{Anti-shock regime for finite-range interactions}
\label{IVd1}

We will see here how we can easily build models with a prescribed maximal width $K$ of the anti-shock and various numbers of longest-lived states in $\mathcal{S}$.

\begin{figure}[ht]
\begin{center}
 \includegraphics[width=0.72\textwidth]{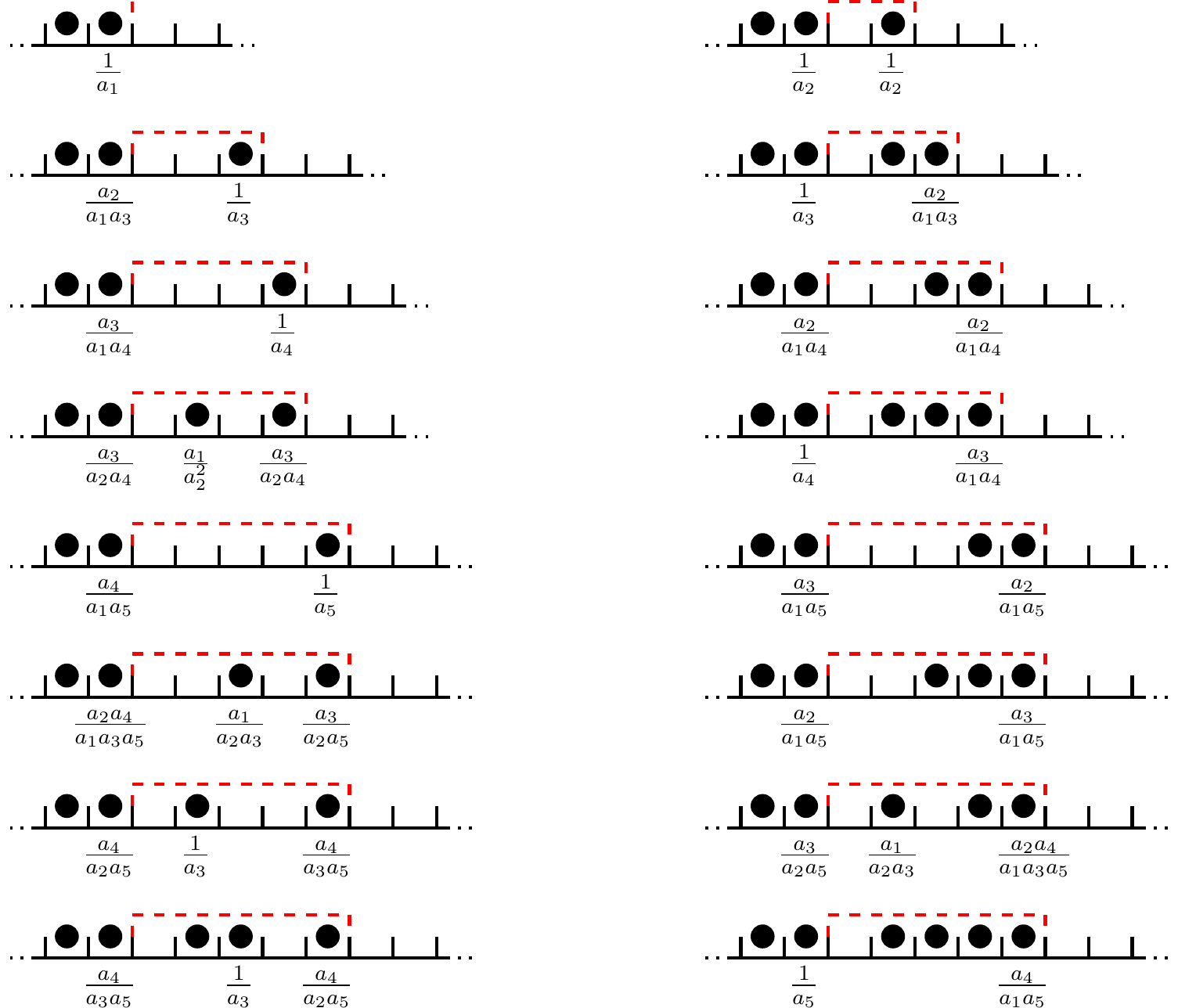}
  \caption{Jump rates from the few simplest anti-shock states.}
\label{fig-LongStates}
 \end{center}
 \end{figure}

We consider a homogeneous system, with $p=1$ except for $p_0$ and $p_L$ at the boundaries, and a symmetric two-bodies interaction $V$ which we will write as
\begin{equation}
V(\{\tau_i\})=-2\sum\limits_{i>j}\log(a_{i-j})\tau_i \tau_j.
\end{equation}
Note that we count every pair of sites $(i,j)$ only once. To make sure that $V$ is short-range, we will take $a_k=1$ for $k>K$. With that notation, the transition rates from anti-shock states in the bulk of the system take a simple form, as seen on fig.\ref{fig-LongStates}: for a configuration with domain walls $10$ at positions $i_k$ and $01$ at positions $j_l$, the jump rate of the particle at position $j$ is given by
\begin{equation}
\frac{\prod\limits_{k}~~a_{i_k-j}}{a_1\prod\limits_{l, j_l\neq j}a_{j_l-j}}.
\end{equation}

~

\begin{figure}[ht]
\begin{center}
 \includegraphics[width=0.72\textwidth]{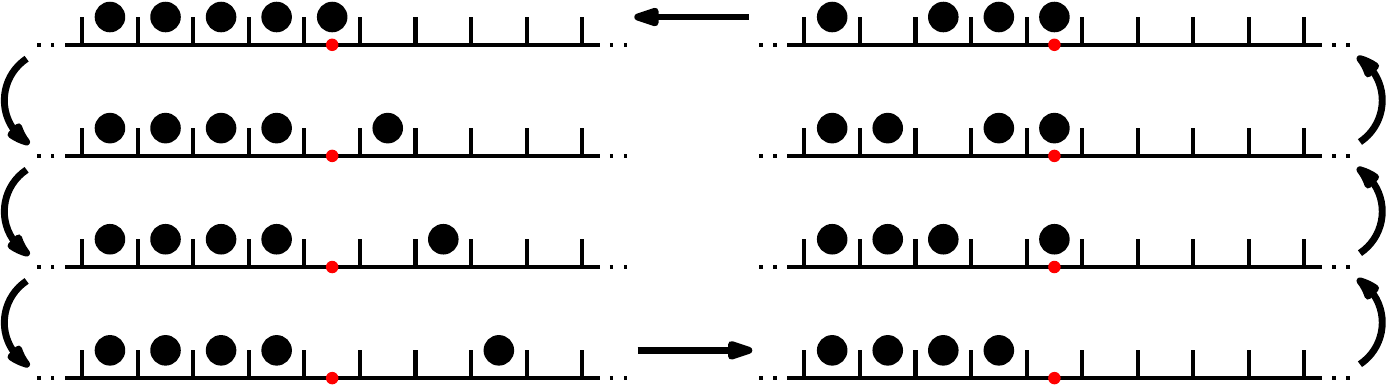}
  \caption{Example of a cycle with $K=4$, $8$ states, and $L+1$ steps. The red dot indicates an arbitrary reference site $k$.}
\label{fig-Cycle}
 \end{center}
 \end{figure}

By choosing $a_k=k$, we get that every escape rate from states with anti-shocks of width less than $K$ is exactly $1$, and all other escape rates are larger than $1$ (the boundary rates also need to be tuned to insure that, which is straightforward). This surprising identity can be easily checked on the examples shown in fig.\ref{fig-LongStates}, and a formal proof can be found in appendix \ref{A2}. This choice of rates yields a number of longest-lived states of order $LK^2/2$. Considering, for instance, the cycle of states shown on fig.\ref{fig-Cycle}, we see that $C<(L+1)/2K$ (as defined in eq.(\ref{mindn})). The full effective process in this case is still quite complicated, so we will not go into more detail.

~

\begin{figure}[ht]
\begin{center}
 \includegraphics[width=0.72\textwidth]{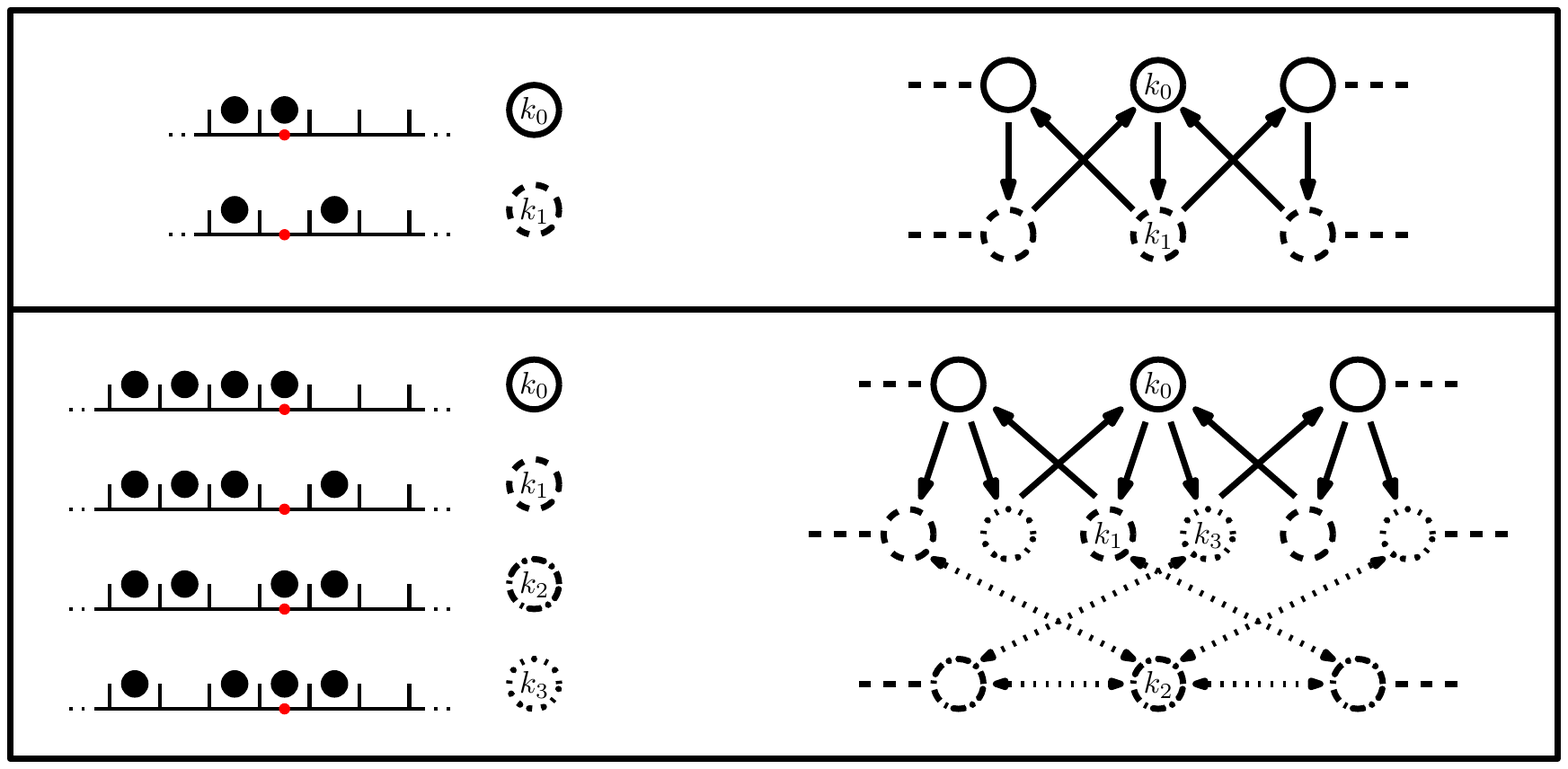}
  \caption{Structure of the effective process for $K=2$ (above) and $K=4$ (below). The red dot indicates an arbitrary reference site $k$. The full arrows indicate transitions which contribute to the effective process at leading order, while the dotted arrows indicate those of lower order, which are not taken into account to estimate the equivalent of $E(\mu)$.}
\label{fig-Effective}
 \end{center}
 \end{figure}

A simpler example is obtained by choosing $a_K=2$ and $a_k=1$ otherwise. In this case, only the anti-shocks of width $0$ or of width $K$ with two possible jumps have an escape rate equal to $1$, all the others being larger. We then have a number of longest-lived states of order $L(K-1)$, and an effective process with a structure given in fig.\ref{fig-Effective}. Note that, except for $K=2$, only three types of anti-shocks end up contributing to the effective dynamics, as all the others only contribute to sub-dominant terms in $\tilde{M}_{eff}$. In all cases, we find that $C=(L+1)/4$, so that $E(\mu)+1\sim {\rm e}^{\mu/4}$ and $g(j)\sim1+4j\log(j)$.

\subsubsection{Non-hydrodynamic behaviour for long-range interactions}
\label{IVd2}

We now exhibit an system with long-range interactions where our result is not valid, to illustrate the importance of that condition. We will construct that example step by step starting from the TASEP with $p_0=p_L=1$, changing one element at a time in order to obtain a model where there is a cycle $\mathcal{O}(L)$ of longest-lived states which makes only $\mathcal{O}(L)$ steps in total.

The simplest such cycle that one could think of is that of one-particle states $|k\rangle=|\{\delta_{i,k}\}\rangle$ plus the empty state $|0\rangle$, where one particle enter from the left, jumps through the whole system, and exits from the right. The escape rates for those states in the case of the TASEP are $1$ from $|0\rangle$ and $|1\rangle$, but $2$ from the other states. Moreover, all perfect (i.e. of width $0$) anti-shock states have an escape-rate of $1$ as well, which we don't want (except for $|0\rangle$ and $|1\rangle$). We need to correct those two issues.

First, we increase the escape-rate of unwanted anti-shocks by adding a repulsive nearest-neighbour interaction
\begin{equation}
V_A(\{\tau_i\})=2\log(A)\sum\limits_{i}\tau_i \tau_{i+1}
\end{equation}
with $A>2$, so that particles leaving a neighbour behind do it with a rate $A>2$, thus disqualifying those states. This has the adverse consequence of facilitating the first jumps from states $|010000...\rangle$ and $|1010000...\rangle$, giving them smaller escape-rates. We correct this by setting $p_1>A$ and $p_2>2-1/A$, so that those states and $|1\rangle$ now have an escape-rate larger than $2$. At this stage, all states $|k>2\rangle$ have an escape-rate of $2$, except for $|0\rangle$ which has an escape-rate of $1$, and all other states have an escape-rate of $2$ or more.

It only remains for us to set the transition rate for $|0\rangle\rightarrow|1\rangle$ to $2$ without modifying the other rates. That is equivalent to adding an interaction potential
\begin{equation}
V_0(\{\tau_i\})=-2\log(2)\prod\limits_{i=1}^{L}(1-\tau_i)
\end{equation}
which is non-zero only for the empty state $|0\rangle$. It is clear that this term does not satisfy the locality condition (\ref{vlocal}), which allows the escape rate from $|L\rangle$ to be the same as that from $|0\rangle$, even though the two possible jumps out of $|L\rangle$ are as far away from each-other as possible and one of them is the same as the jump out of $|0\rangle$.

\begin{figure}[ht]
\begin{center}
 \includegraphics[width=0.5\textwidth]{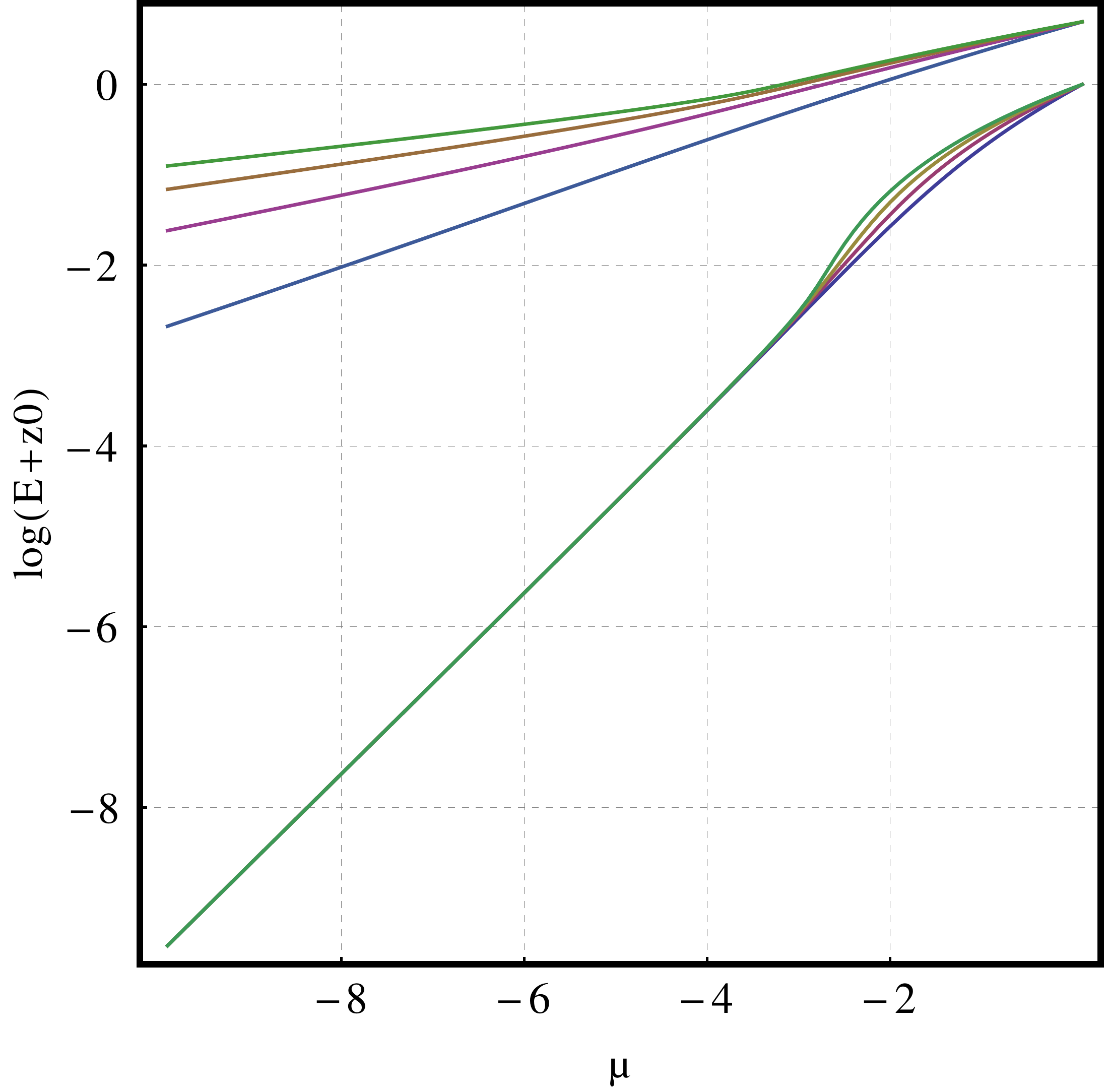}
  \caption{Numerical evaluations of $\log(E(\mu)+z_0)$ with $V_0$ (top curves) and without $V_0$ (bottom curves), for various system sizes: from bottom to top, $L=4$ (blue), $L=6$ (purple), $L=8$ (orange), $L=10$ (green).}
\label{fig-LogE}
 \end{center}
 \end{figure}

It follows from that cycle of states that $C=(L+1)/(L-1)$, as defined in eq.(\ref{mindn}), and that $E(\mu)+z_0\sim {\rm e}^{\mu/(L-1)}$, so that 
\begin{equation}\boxed{
g(j)\sim 2+Lj\log(j).
}\end{equation}
The factor $L$ means that the probability of observing a small current scales with the size of the system, as expected. This can be seen clearly on numerical evaluations of $\log(E(\mu)+z_0)$ for small system sizes. On fig.\ref{fig-LogE}, we plot that quantity for systems of various sizes and with or without the non-local interaction $V_0$. As we can see, it is independent of $L$ even for small negative values of $\mu$ when $V_0$ is not introduced, and depends strongly on $L$ when it is.

\newpage

\section{Illustration and discussion}
\label{V}

In this final section, we illustrate our results with a variety of numerical plots, and discuss a possible physical interpretation of the dynamical transition in terms of maximum hydrodynamic current as well as a possible connexion to the KPZ universality class.

Considering the broad class of models that we have been looking at, constrained only by (\ref{pbound}), (\ref{vbound}) and (\ref{vlocal}), we need to somewhat restrict ourselves in order to obtain something meaningful in the large size limit (i.e. an identifiable phase transition). Above all, that limit itself has to make sense, which restricts the sequence of models of increasing size (or possibly the sequence of ensembles of models of increasing size) that we may consider. We first have to distinguish between disordered models, for which $p_i$ and/or $V$ might be drawn from a distribution, and models with fixed parameters. We then have to define $p_i$ and $V$ so that a large size limit can be taken, which might involve taking $p_i$ to be a discretisation of a fixed smooth function $p(x)$, and $V$ to be a combination of simple short- or finite-range $n-$body interactions, with perhaps a slow space-dependence.

In this section, we will only be considering (and conjecturing about) homogeneous systems with simple short-range potentials, of which the ASEP is the simplest example (and we will use it as a guide throughout, with all the related results taken from \cite{Lazarescu2015}).

~

 \begin{figure}[ht]
\begin{center}
 \includegraphics[width=0.6\textwidth]{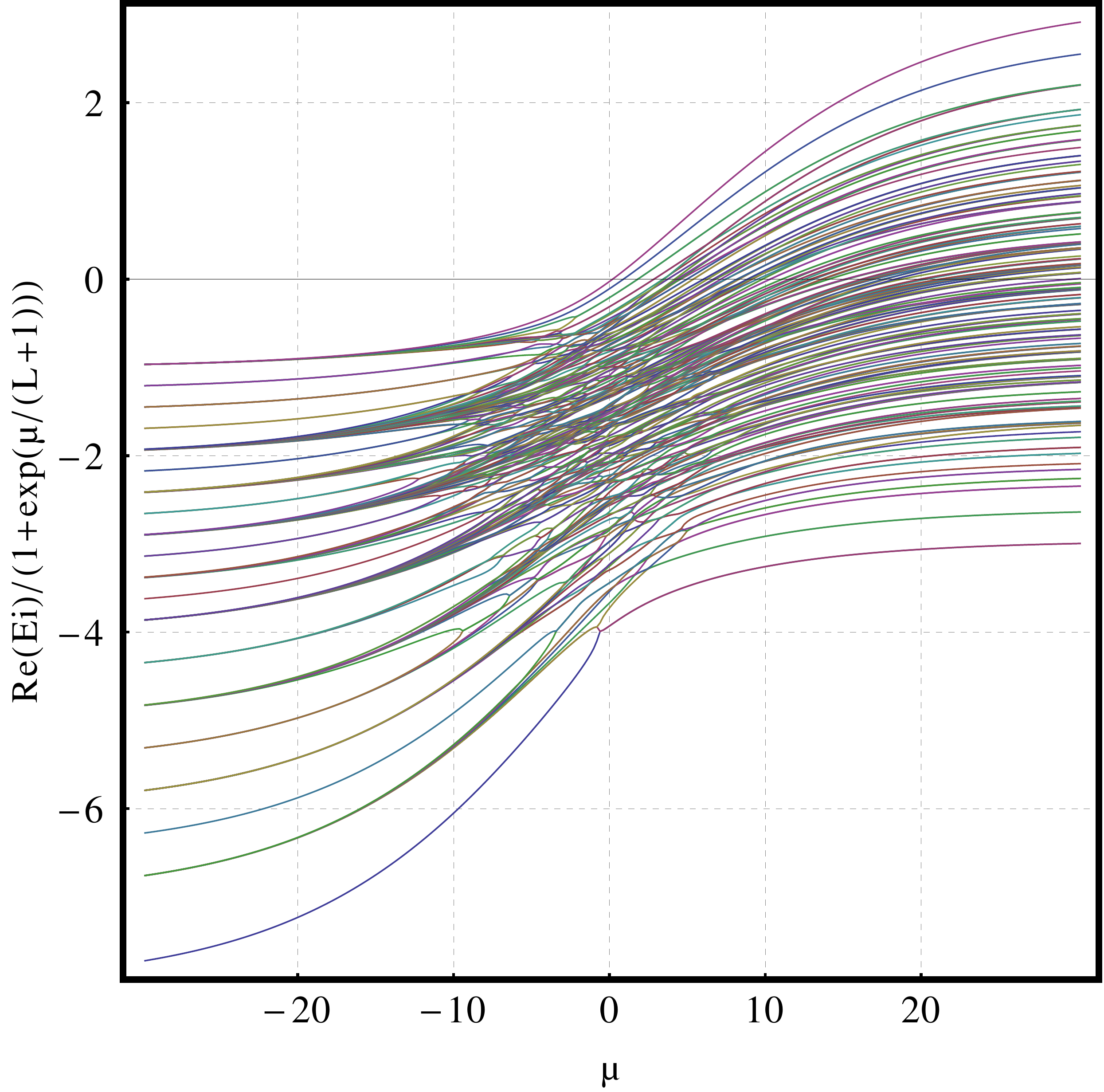}
  \caption{Rescaled real part of the spectrum of $M_{\mu}$ for a model of size $L=8$, with homogeneous jumps $p_i=1$ and next-to-nearest-neighbour interactions $a_2=2$. The colours are only there to help differentiate overlapping curves.}
\label{fig-E3}
 \end{center}
 \end{figure}
\begin{figure}[ht]
\begin{center}
 \includegraphics[width=0.8\textwidth]{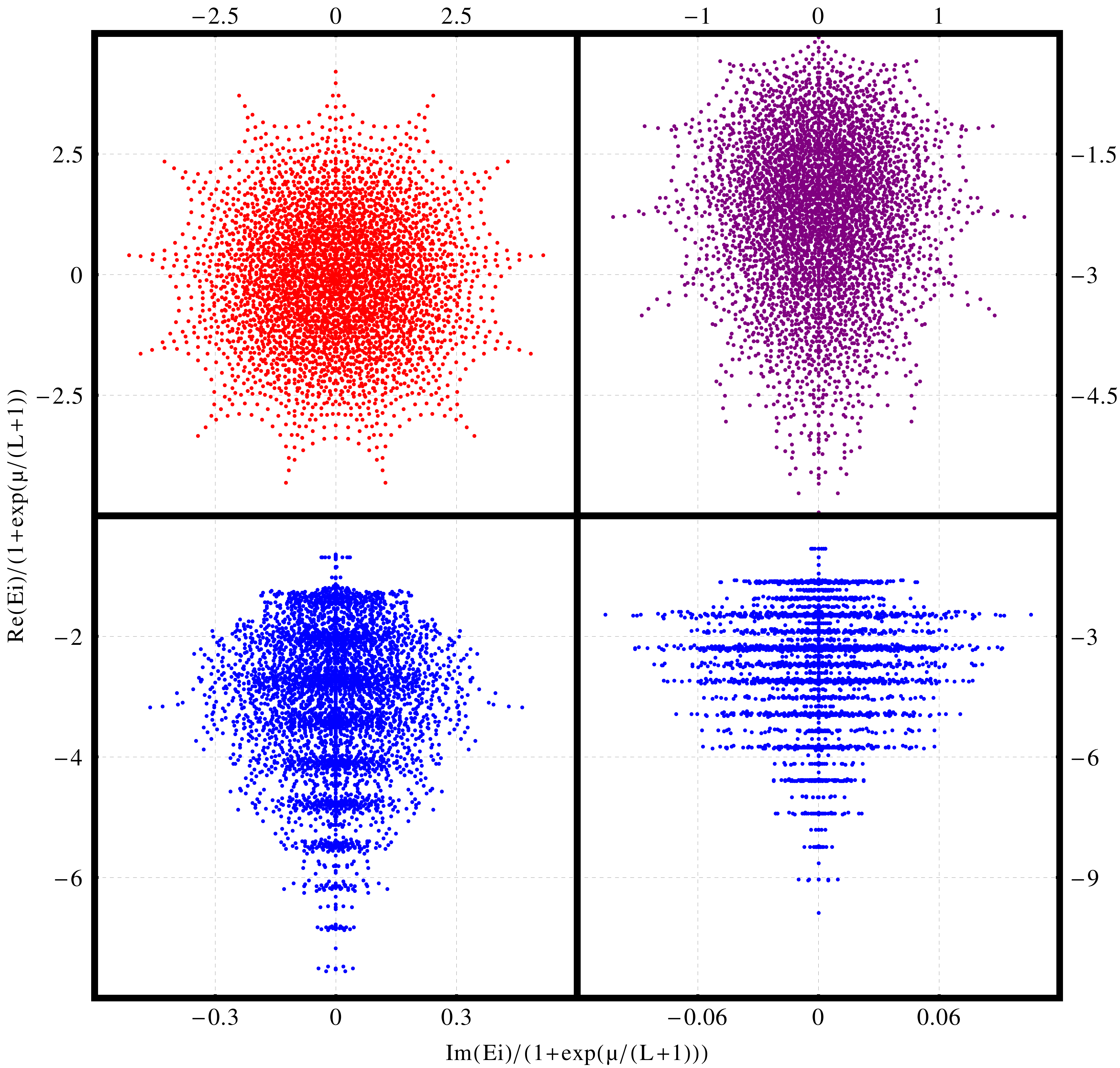}
  \caption{Rescaled complex spectra of $M_{\mu}$ for a model of size $L=12$, with homogeneous jumps $p_i=1$ and nearest-neighbour interactions $a_2=2$, at various values of $\mu$: $50$ (red, top left), $0$ (purple, top right), $-10$ (blue, bottom left) and $-20$ (blue, bottom right). Note that each plot has a different scale.}
\label{fig-EEEE}
 \end{center}
 \end{figure}

We start by simply looking at the qualitative behaviour of the spectrum $\{E_i\}$ of $M_\mu$ as a function of $\mu$. For this, we choose a simple case where all $p_i=1$, with next-to-nearest-neighbour interactions $a_2=2$ in the notation of section \ref{IVd1}, and a system size $L=8$. We plot, on fig.\ref{fig-E3}, the real part of the eigenvalues of $M_\mu/(1+{\rm e}^{\mu/(L+1)})$, where the rescaling is introduced so that those eigenvalues converge to a constant for $\mu\rightarrow\infty$ rather than diverge, which makes the plot clearer. We also plot on fig.\ref{fig-EEEE} the rescaled complex spectrum of the same model with $L=12$ for a few values of $\mu$.
 
As we can clearly see, the aspect of the spectrum is quite different between negative and positive values of $\mu$. For $\mu\rightarrow\infty$, we have a Fermionic spectrum, where the eigenvalues are distributed according to the semi-circle law (as is clear on fig.\ref{fig-EEEE} ; the highest and lowest eigenvalues on fig.\ref{fig-E3} seem to be separated from the rest of the spectrum by a rather large gap, but this is due to the small system size and would not be the case in the large size limit). In the $\mu\rightarrow-\infty$ limit, on the other hand, we have a quasi-discrete real part of the spectrum where eigenvalues accumulate around specific values with a high degeneracy (this is helped by the fact that we chose an homogeneous system with a simple next-to-nearest-neighbour potential, in order to have a high degeneracy of escape rates even for a small size~; in the large size limit, we would observe such a high degeneracy even for more complex potentials and slowly-varying jump rates). The transition between these two regimes is clearly visible on fig.\ref{fig-E3} as an area dense with bifurcations and crossings, extending between $\mu\sim-10$ and $\mu\sim8$, and is an indication of the potential existence of a phase transition in the large size limit, even though for such a small size no sign of a non-analiticity can be yet observed for the highest eigenvalue $E(\mu)$. It is unclear how that area itself behaves in the large size limit, but it would be reasonable to expect that it converge to a single non-analiticity at $\mu=0$ under the proper rescaling.

~

That behaviour is similar to that of the open ASEP, as analysed in \cite{Bodineau2006,Lazarescu2015}. In the low current regime, we have a \textit{hydrodynamic} phase where the large deviation function of the current is consistent with applying the macroscopic fluctuation theory (MFT, \cite{Bertini2007}) or the additivity principle \cite{PhysRevLett.92.180601} to the continuous limit of the model, with a diffusion constant of order $L^{-1}$, which is to say that the long time large deviation function $g(j,\rho)$ of the current $j$ and mean local density $\rho(x)=\langle\tau_{\lfloor xL\rfloor}\rangle$ has the form
\begin{equation}\label{gjr}
g(j,\rho)=L\int_0^1\frac{\bigl[j-J^\star(\rho)+\frac{D(\rho)}{L}\nabla\rho\bigr]^2}{2\sigma(\rho)}dx
\end{equation}
with boundary conditions $\rho(0)=\rho_a$ and $\rho(1)=\rho_b$ and can be minimised over $\rho$ in order to obtain $g(j)$. In the case of the ASEP, we have that $\sigma(\rho)=\rho(1-\rho)$, $D=\frac{p}{2}$ and $J^\star(\rho)=p~\sigma(\rho)$ (this proportionality being a far-from-equilibrium version of Einstein's relation, which might be specific to the ASEP and a few other models and is not to be expected in general). Moreover, the minimisation produces not only the most probable density $\rho^\star$, which is associated to $E(\mu)$ through the Legendre transform of $g(j,\rho^\star)$, but also a family of metastable states $\rho_i$, which turn out to be related to the other eigenvalues $E_i(\mu)$ in the same way (c.f. \cite{Lazarescu} for more details). The structure of $\{E_i(\mu)\}$ thus obtained is similar to what is observed on fig.\ref{fig-E3} for $\mu$ low enough. The first group of eigenvalues then account for the effective process discussed in section \ref{IVa} at lowest order, and subsequent groups account for higher orders. Moreover, those eigenvalues undergo many first-order phase transitions as the boundary parameters $\rho_a$ and $\rho_b$ are varied.

In the high current regime, as we have seen, the behaviour of the ASEP is exactly the same as that of the inhomogeneous interacting versions that we have considered: we find a \textit{correlated} free Fermion phase characterised by an average density of $\frac{1}{2}$ with long-range correlations, which is sometimes called a hyperuniform phase \cite{Jack2015}, and which can be described by a conformal field theory \cite{Karevski2016}.

Between those two regimes is a dynamical phase transition, which occurs when the fluctuating current $j$ goes through the critical value $j=\frac{p}{4}$. That transition is of second order, and can be observed directly on the minimisation of eq.(\ref{gjr}), since a value $j>\frac{p}{4}$ yields a minimum of order $L$ (the integrand is finite for every $x$) whereas for $j<\frac{p}{4}$ it is of order $1$ (the integrand is non-zero only on a fraction of order $L^{-1}$ of $[0,1]$). However, the behaviour of $g(j)$ for $j\rightarrow \frac{p}{4}^+$ is wrongly predicted by the MFT as $(j-\frac{p}{4})^2$, whereas exact calculations using integrability methods \cite{Lazarescu2014,Lazarescu2015} yield instead a true scaling as $(j-\frac{p}{4})^{\frac{5}{2}}$.
 \begin{figure}[ht]
\begin{center}
 \includegraphics[width=0.8\textwidth]{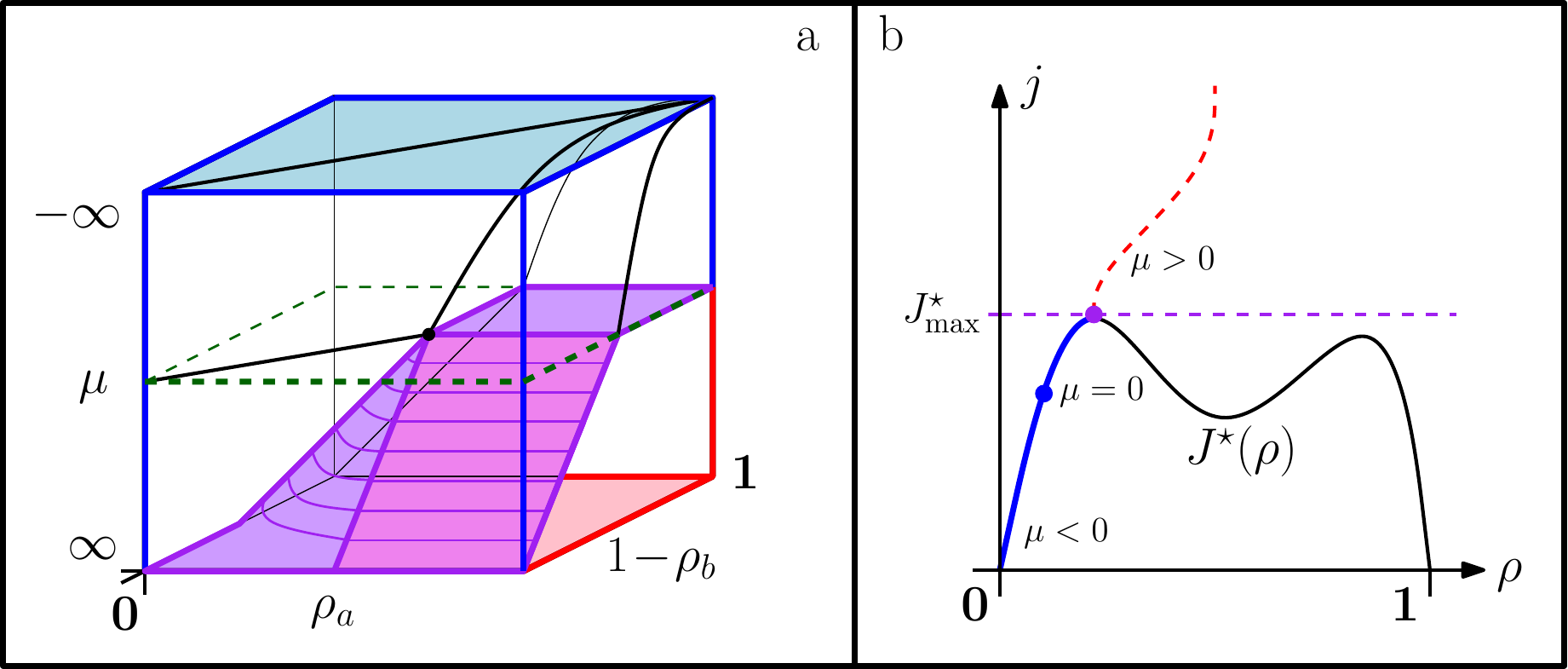}
  \caption{a) Dynamical phase diagram of the current for the open ASEP, with respect to the boundary densities $\rho_a$ and $\rho_b$, and the biasing parameter $\mu$. The green dashed square corresponds to the static phase diagram of the current in the stationary state $\mu=0$. The hydrodynamic regime (top) is highlighted in blue and the correlated one (bottom) in red, with the second-order dynamical phase transition in between in purple. \newline b) Density/current diagram for a hypothetical model with finite-range interactions. The black curve $J^\star(\rho)$ corresponds to the hydrodynamic (mean-field) current associated to a uniform density $\rho$. An example of a trajectory $(j(\mu),\rho(\mu))$ for fixed boundary densities is represented with colours corresponding to those on the phase diagram.}
\label{fig-PhaseSigma}
 \end{center}
 \end{figure}

The dynamical phase diagram of $E(\mu)$ is shown on fig.\ref{fig-PhaseSigma}.a, as a function of the boundary parameters $\rho_a$ and $\rho_b$. The purple surface corresponds to the hydrodynamic/correlated phase transition (the distinction between the light and dark purple zones will be made later). The region above, marked in blue, is the hydrodynamic phase corresponding to $j<\frac{p}{4}$ and contains a variety of first-order phase transitions between different typical states $\rho^\star$. The region below, marked in red, is the correlated phase, and does not contain other phase transitions as far as we know, although a precise description of the bulk of the phase is yet to be obtained.

~

In light of these similarities, we are led to give the same interpretation to the physical origin of the dynamical phase transition as in \cite{Lazarescu2015} for the ASEP: for low enough currents, the system behaves, in the large size limit, in accordance with a Langevin equation with conserved noise
\begin{equation}\label{LangevinGas}
{\rm d}_t \rho=-\nabla j~~~~{\rm with}~~~~j=J^\star(\rho)-\frac{D(\rho)}{L}\nabla\rho+\sqrt{\sigma(\rho)}\xi,
\end{equation}
where $\xi$ is a Gaussian white noise. The so-called transport coefficients $J^\star(\rho)$, $D(\rho)$ and $\sigma(\rho)$ are in general very difficult to obtain from the microscopic process \cite{Arita2014,Arita2016}, but one important property of $J^\star(\rho)$ can be deduced from the exclusion property alone: it vanishes at $\rho=0$ and at $\rho=1$, since those densities cannot sustain any current, which means that $J^\star(\rho)$ is bounded from above, with a maximum value $J^\star_{\mathrm{max}}$.

For a fluctuating current smaller than that maximum (as well as for the steady state, which lies somewhere along $j=J^\star(\rho)$), there are hydrodynamic states which produce it through localised defects at a cost $g(j)$ which doesn't grow with $L$. In the limit of very low currents, for instance, the typical densities are $0$ and $1$, and the localised defects in question are the anti-shocks that bridge those two densities.

For a current larger than that maximum, the best hydrodynamic states have a cost proportional to $L$, but it becomes more efficient to introduce correlations everywhere in the system, producing a hyperuniform state which is optimised to produce a large current by forcing the particles and holes to alternate more than at random. The dynamical phase transition therefore corresponds to the appearance of correlations in the system when the current is pushed beyond its hydrodynamic regime.

This is illustrated schematically on fig.\ref{fig-PhaseSigma}.b, where a hypothetical $J^\star$ is represented (as would appear for instance with nearest-neighbour interactions, in a KLS-type model \cite{PhysRevB.28.1655,Popkov1999}), along with a trajectory $(j(\mu),\rho(\mu))$ obtained by varying $\mu$ from $-\infty$ to $\infty$ for certain fixed boundary densities $\rho_a$ and $\rho_b$ (which is to say a vertical line from the diagram on fig.\ref{fig-PhaseSigma}.a): the blue curve represents the fluctuating hydrodynamic regime, from a completely empty state $\rho=0$ to a state (purple dot) sustaining the maximal hydrodynamic current (purple dashed line), passing through the typical stationary state (blue dot) at $\mu=0$ ; the red dotted curve corresponds to the correlated regime for currents larger than that maximum.

A more recent study \cite{Baek2016} makes a very similar conjecture for boundary-driven systems, consistent with ours: for systems obeying Einstein's relation $J^\star\propto D\sigma$, one can observe a hydrodynamic behaviour for currents even far from equilibrium, as long as they are lower than the maximum of $J^\star$ if that maximum exists. When $J^\star$ is unbounded, as for some zero-range processes, no such dynamical phase transition can be found.

~

Finally, we should note that the hydrodynamic/correlated dynamical phase transition for the ASEP is closely related to the appearance of Tracy-Widom distributions in the statistics of the relaxation of $j$ towards stationarity, as seen in models from the KPZ universality class \cite{Spohn2016,Corwin2016}. The context there is quite different: the models are of infinite size and observed at long times, whereas we put ourselves at infinite time and then increase the size. However, it would make sense that both approaches be at least somewhat connected, and a first confirmation of this is the fact that the phase diagrams of small deviations around the typical currents are identical (although the current in question and the two parameters are not exactly the same in both cases).

More precisely, we compare:
\begin{itemize}
\item 1) the vicinity of the central slice (marked by a green dashed line) of fig.\ref{fig-PhaseSigma}.a, which corresponds to small fluctuations of the time-averaged stationary current of the open ASEP with boundary densities $\rho_a$ and $\rho_b$, in the large size limit (we refer to section VI.B.5 of \cite{Lazarescu2015} for the names and full description of the phases) ;
\item 2) fig.2 from \cite{BenArous2011}, which corresponds to the fluctuations of the time-integrated current $tj$ across the middle bond of an ASEP on an infinite line with a product state initial condition with densities $\rho_a$ on the left ($i<0$) and $\rho_b$ on the right ($i>0$), in the large time $t$ limit, on a certain time-scale (which is $t^{\frac{1}{2}}$ outside of the maximal current phase $(\rho_a>\frac{1}{2},\rho_b<\frac{1}{2})$, and $t^{\frac{1}{3}}$ inside of it, including the boundaries).
\end{itemize}

The comparison is quite straightforward:
\begin{itemize}
\item the parts of 1) which do not sit at a phase transition, i.e. the LD  phase $(\rho_a<\frac{1}{2},1-\rho_b>\rho_a)$ and HD phase $(1-\rho_b<\frac{1}{2},\rho_a>1-\rho_b$), correspond to Gaussian fluctuations on a diffusive time-scale $t^{\frac{1}{2}}$ in 2) ;
\item the line which corresponds to a first order phase transition between two states in 1), i.e. the S line $(\rho_a<\frac{1}{2},1-\rho_b=\rho_a)$, corresponds to the maximum of two Gaussian distributions on a diffusive time-scale $t^{\frac{1}{2}}$ in 2) ;
\item most importantly, the parts of 1) which sit at the hydrodynamic/correlated dynamical phase transition, i.e. the MC phase $(\rho_a>\frac{1}{2},1-\rho_b<\frac{1}{2})$ and its boundaries, correspond to three different Tracy-Widom distributions on a subdiffusive time-scale $t^{\frac{1}{3}}$ in 2).
\end{itemize}

On this last point, in both cases, the special statistics of the current arise from the fact that it is much more difficult for the system to accommodate for currents higher than average than for lower ones. It is then natural to wonder which features of these statistics are universal, and which are model-dependent, or even parameter-dependent within the same model. Considering even the standard ASEP, the transition surface naturally splits in four areas, corresponding to the four sectors $\rho_{\{a,b\}}\lessgtr\frac{1}{2}$, of which we have mentioned only one so far, namely $(\rho_a>\frac{1}{2},\rho_b<\frac{1}{2})$. It turns out that, because of a very special symmetry of the model (proven in appendix \ref{A3}, extending on a result from \cite{Torkaman2015}), the sector $(\rho_a<\frac{1}{2},\rho_b>\frac{1}{2})$ has the same properties. On the other hand, the line $(\rho_a=\frac{1}{2},\rho_b>\frac{1}{2})$ (which is, mysteriously, equivalent to a half-filled periodic system \cite{derrida1999universal}), on the boundary of the MC phase, and the point $(\rho_a=\frac{1}{2},\rho_b=\frac{1}{2})$, have slightly different properties (same exponents but different pre-factors and distributions), and we expect the sector $(\rho_a>\frac{1}{2},\rho_b>\frac{1}{2})$ and its symmetric to make for yet another universality subclass. 

All that being said, we expect models with more complex hydrodynamic currents $J^\star$ to all be in the same universality class. However, situations where several densities produce the same maximal current (for instance a version of fig.\ref{fig-PhaseSigma}.b with a symmetric $J^\star$) would most probably show different behaviours in the appropriate regimes, and there is undoubtedly much more to be understood about the hydrodynamic/correlated dynamical phase transition. In particular, we should be able to find some correspondence between the exponents and pre-factors found around the transition in the stationary case and those obtained in the infinite volume case from the perspective of so-called third order phase transitions \cite{Majumdar2014,LeDoussal2016}, which we believe to be the long-time relaxation equivalent to our stationary dynamical phase transition.

\newpage

\section{Conclusion}

In this paper, we have analysed the large deviations of the current $g(j)$ in extreme limits for a very general class of models based on the TASEP, with inhomogeneities and short-range interactions. After defining the models and formalism relevant to our endeavour, we reduced the problem to that of obtaining the approximate behaviour of the largest eigenvalue $E(\mu)$ of the Markov matrix deformed by a counting parameter $\mu$, in the limits of $\mu\rightarrow\pm\infty$.

In the $\mu\rightarrow\infty$ limit, corresponding to a high current, the deformed Markov matrix is equivalent to a free Fermions Hamiltonian and depends only trivially on the disorder and interactions. We found that $g(j)$ is proportional to the size of the system $L$ and that the typical states are Coulomb gases, with an average density equal to $\frac{1}{2}$ and strong nearest-neighbour anti-correlations, resulting in a \textit{correlated} phase. In the $\mu\rightarrow -\infty$ limit, corresponding to a low current, the deformed Markov matrix is a high-order perturbation of a diagonal matrix. We were able to show that the typical states are of the anti-shock type, with a block of particles followed by a block of holes, separated by an area no larger than a certain constant, and in particular not growing with the system size $L$. Moreover, $g(j)$ was also shown not to scale with $L$, consistently with being in a \textit{hydrodynamic} phase. Those two very different limits, and in particular the different scaling of $g(j)$ with respect to $L$, indicating the possible existence of a dynamical phase transition in between.

We then looked at a specific model for illustration, with homogeneous rates and next-to-nearest-neighbour interactions. We saw how the transition between a hydrodynamic regime and a correlated one manifested itself on the whole spectrum of the deformed Markov matrix, and interpreted that transition in terms of $j$ pushing beyond the maximal hydrodynamic current $J^\star_{\mathrm{max}}$ at the price of introducing correlations in the system. We also showed a connexion between that dynamical phase transition and the appearance of Tracy-Widom distributions, as for all models in the KPZ universality class, in the infinite volume case.

~~

These results are a quite encouraging step towards understanding the large size limit of interacting particle models far from equilibrium, especially because of the non-solvable nature of the models we have considered: the behaviour that we have described comes only from the geometric structure of the models (i.e. a lattice gas, with physical inhomogeneities and interactions), and not from a very special algebraic structure of the Markov matrix. This makes it likely that our methods could be applied for many other models with different components or geometries (it would for instance be quite straightforward to apply them to the totally asymmetric partial exclusion process, where the number of particle per site is not limited to one but to some integer \cite{Arita2014}, or to a multispecies TASEP \cite{Crampe2016,Crampe}). Of course, we were only able to perform calculations in extreme limits where the problem is greatly simplified, and it is more than likely that we will have to rely at least partly on numerics if we want to go further, for instance in describing the dynamical transition itself rather than the phases on each side of it. Luckily, a lot of progress is being made on the variety and effectiveness of the numerical methods available for that purpose \cite{Gorissen2009,giardina2011simulating,Espigares2013,Nemoto2016}.

There is of course a lot more to be done on the subject. One of the outstanding problems in this context, for instance, is to obtain the hydrodynamic transport coefficients, as seen in eq.(\ref{LangevinGas}), from the microscopic dynamics of the systems, even in an equilibrium setting \cite{Arita2014,Arita2016}. Once those coefficients are known, and the MFT is assumed to be valid, one can analyse the full spectrum of the model in the bulk-driven case, and identify the effective dynamics to any order (this will be the subject of a future work \cite{Lazarescu}), as well as relate them to the relaxation paths and the pseudopotential which have already been studied \cite{Bahadoran2010}. On the other side of the transition, a good description of the correlated phase remains to be found except in the infinite current limit \cite{Karevski2016}, and in particular the appropriate order parameters have not been clearly identified, although it seems likely that they would consist of correlation functions (as the local density becomes irrelevant in that phase, and the correlations grow from $0$ to a finite value). As for the transition itself, we have already mentioned that different situations will give rise to different pre-factors to the scaling of $g(j)$, related to different sub-classes within KPZ universality, and a precise classification of those does not exist yet as far as we know. Finally, we have focused on models with well-behaved jump rates and potentials, but our general result holds for disordered systems as well, although the consequences are more mysterious in that case and remain to be analysed. It is for instance unclear under which conditions we can expect a hydrodynamic phase to survive, as it does for dilute disorder \cite{Bahadoran2016}.

~

We conclude by examining in more detail a few of the natural extensions of our results as well as how they relate to other existing works.
\begin{itemize}
\item \textit{Partially asymmetric models}: In order to take the $j\rightarrow0$ limit through $\mu\rightarrow-\infty$, we had to restrict ourselves to totally asymmetric models. It is quite likely that the phenomenology of partially asymmetric models would be exactly the same, as was shown for the ASEP in \cite{Lazarescu2015}, where the only difference is a simple factor $(p-q)$ which rescales the current (meaning that, by some miracle, Einstein's relation $J^\star\propto D\sigma$ is valid even far from equilibrium), although in general we would expect $J^\star$ to depend on the backwards rates in a more complicated way. However, extending our method to that case does not seem easy: first of all, the expansion around $j=0$ now has to be done at a finite value of $\mu$, which means that the deformed Markov matrix is not a perturbation of a diagonal matrix any more ; and secondly, the possibility of backward jumps makes eq.(\ref{MeffPath}) much more complex, since every term now contains an infinity of paths that needs to be re-summed.
\item \textit{Close-to-equilibrium systems}: The case of boundary-driven or weakly bulk-driven systems is more significantly different. Unlike the asymmetric case, the MFT is rigorously proven for low currents (note that in that case, the stationary current is itself ``low", as it is of order $L^{-1}$ if we measure it on one bond only, and it is the space-integrated current which converges to a finite value), but the typical states are not discontinuous, as there is no difference in scaling between the drift and diffusion terms in the MFT equation equivalent to eq.(\ref{LangevinGas}). On the other hand, the large current limit is exactly the same as here since the $\mu\rightarrow\infty$ limit is identical. We therefore expect to find a dynamical phase transition, and we already mentioned consistent recent results for large fluctuating currents in close-to-equilibrium models \cite{Baek2016}. However, the nature of the transition might well be different : in periodic systems, a dynamical phase transition has been identified \cite{PhysRevE.72.066110,Bodineau2008,appert2008universal,Simon2011}, in which travelling waves states seem to play an important role \cite{PhysRevE.72.066110,Espigares2013}, although that might be an effect of the total density constraint in periodic systems, as no such states are found in open systems \cite{Lecomte2010,Baek2016a}. Moreover, the appropriate scaling for the current is not the same as in the bulk-driven case (i.e. the space-integrated current is finite, not the one-bond current), which means that the high current part of the large deviation function loses its scaling with respect to $L$.
\item \textit{Large deviations of the activity}: Another quite natural extension would be to consider observables other than the current, such as for instance the dynamical activity, defined as a symmetric average of the number of jumps in both directions, rather than an antisymmetric one. Note that, unlike the current, the activity is not conserved throughout the system, and so every different weighting of the single bond activities is a different observable, the standard one being the uniform average (or unweighted sum). For totally asymmetric models, the current and average activity are one and the same, but they play different roles close to equilibrium (as the current appears explicitly in the MFT action, and the activity does not). The activity is known to undergo dynamical phase transitions for exclusion processes \cite{Bodineau2008,Lecomte2012,Jack2015} as well as for kinetically constrained models \cite{Garrahan2009,Nemoto2016,Nemoto2014}. For the former, the transition must be somewhat linked to that of the current, since they are identical in the totally asymmetric limit, although one should note that the choice of scaling of the observable (total activity or average activity, the latter being divided by the system size $L$) will have an effect on the aspect of the phase transition, which will appear to be of first order for the total activity, as it would for the total current \cite{Jack2015,1751-8121-44-11-115005}. For more general models, or more general definitions of the activity, we expect our methods to be applicable, although perhaps not as straightforwardly as for the current. In the high activity limit, the deformed Markov matrix would be equivalent to an inhomogeneous XX spin chain, solvable in principle, and the same scaling would be found for the large deviation function. In the low activity limit, we would still have a perturbation around the diagonal, but with backward jumps allowed, which seems to result in $\varepsilon^2$ terms being present in $E(\mu)$, meaning that the large deviation function of the activity would scale linearly in $L$ (excluding the constant part) as it does in the high activity regime. That would invalidate our scaling argument for the existence of a dynamical phase transition. A more careful analysis is in order.
\item \textit{Higher dimensions}: Finally, we might wonder if our methods can be extended to other geometries. The first step would be to consider the model on a tree, where it is known that the stationary density to current relation is consistent with a hydrodynamic behaviour at least in some cases \cite{Mottishaw2013}. We expect to be able to extend the low current method without major issues. In the high current limit, although the standard free Fermion techniques we used here are expected to fail, one might still be able to perform calculations using auxiliary spins at every fork, as done in \cite{Crampe2013} for a star graph. In the case of higher-dimensional regular lattices, things might be less straightforward, as there are more ways to be far for equilibrium than in one dimension: the system can be driven along loops rather than from one side to the opposite one. Moreover, the stationary current itself will be more complex than in one dimension, as the zero divergence condition allows for vortices in addition to a constant overall flux. In the relatively simple case of a system driven along one of the lattice directions, between reservoirs, with periodic or closed boundary conditions in the other directions, we expect our methods to be applicable but to produce highly degenerate dominant states at leading order, which then have to be separated by a perturbation to a higher order (especially in the high current limit, where the system essentially splits into disconnected one-dimensional chains at leading order). The low current limit can then be compared to the MFT approach, where a dynamical phase transition has already been found close to equilibrium \cite{Hurtado2011}.
\end{itemize}

~~

~~

\textit{Acknowledgements}: I would like to thank M. Esposito and his group, as well as C. Maes, G. Schutz and D. Karevski, for interesting and useful discussions. I am grateful to R. Jack for helping me correct a mistake in eq.(\ref{LowCurrtEquiv}). This work was supported by the Interuniversity Attraction Pole - Phase VII/18 (Dynamics, Geometry
and Statistical Physics) at KU Leuven and the AFR PDR 2014-2 Grant No. 9202381 at the University of Luxembourg.

\newpage

\appendix

\section{Asymptotics of the Legendre transform}
\label{A1}

In this appendix, we check that the Legendre transform of the asymptotic equivalent of $E(\mu)$ is an asymptotic equivalent of the Legendre transform of the real $E(\mu)$.

Fist of all, we may note that this is far from being guaranteed. Consider for instance:
\begin{equation}
{\rm e}^{x}\xrightarrow[{\rm Legendre}]{}y\log(y)-y~~\sim~~ y\log(y)\xrightarrow[{\rm Legendre}]{}{\rm e}^{x-1}
\end{equation}
for $x\rightarrow\infty$ and $y\rightarrow\infty$, where $a\sim b$ means $a=b+o(b)$. The functions ${\rm e}^{x}$ and ${\rm e}^{x-1}$ are not equivalent to each other, even though their Legendre transforms are. The issue comes from ${\rm e}^{x}$ not being algebraic.

Let us now focus on $E(\mu)$ in particular. It is the largest eigenvalue of a matrix whose entries are algebraic functions of $x={\rm e}^{\mu/(L+1)}$ (either proportional or inversely proportional to it, or constant), so it is itself an algebraic function $E(\mu)=F(x)$ of $x$. We need to consider both $x\rightarrow\infty$ and $x\rightarrow 0$.

~~

Consider first $x\rightarrow\infty$ (i.e. $j\rightarrow\infty$) and examine the large $x$ asymptotics of $F$: since it is algebraic, we have $F(x)= Ax^{\alpha}+\mathcal{O}(x^{\alpha-1})$ for some constants $A$ and $\alpha$. We also have that $F'(x)= A\alpha x^{\alpha-1}+\mathcal{O}(x^{\alpha-2})$, and more importantly that the inverses of these functions are also algebraic, with for instance $\bigl[xF'(x)\bigr]^{-1}(y)=(y/A\alpha)^{1/\alpha}+\mathcal{O}(y^{1/\alpha-1})$. The Legendre transform of $E(\mu)$ is then
\begin{equation}
g(j)=j\mu-E(\mu)~~~~~~\mathrm{with}~~~~~~ j=E'(\mu)=\frac{x}{L+1}F'(x)= \frac{A\alpha}{L+1} x^{\alpha}+\mathcal{O}(x^{\alpha-1}),
\end{equation}
so that
\begin{align}
x&= \biggl(\frac{(L+1)j}{A\alpha}\biggr)^{1/\alpha}+\mathcal{O}(j^{1/\alpha-1})\\
\mu&=\frac{L+1}{\alpha}\log\biggl(\frac{(L+1)j}{A\alpha}\biggr)+\mathcal{O}(j^{-1})\\
E(\mu)&=\frac{(L+1)j}{A\alpha}+\mathcal{O}(1)
\end{align}
yielding finally
\begin{equation}
g(j)= \frac{(L+1)}{\alpha}\Biggl(j\log\biggl(\frac{(L+1)j}{A\alpha}\biggr)-j\Biggr)+\mathcal{O}(1)
\end{equation}
so that the Legendre transform of $A{\rm e}^{\alpha\mu/(L+1)}\sim E(\mu)$ is indeed an equivalent of $g(j)$.

~~

Consider now $x\rightarrow 0$ (i.e. $j\rightarrow 0$) and $F=-z_0+ Ax^{\alpha}+\mathcal{O}(x^{\alpha+1})$ for some constants $A$ and $\alpha$. We also have that $F'(x)= A\alpha x^{\alpha-1}+\mathcal{O}(x^{\alpha})$, and $\bigl[xF'(x)\bigr]^{-1}(y)=(y/A\alpha)^{1/\alpha}+\mathcal{O}(y^{1/\alpha+1})$. The Legendre transform of $E(\mu)$ is then
\begin{equation}
g(j)=j\mu-E(\mu)~~~~~~\mathrm{with}~~~~~~ j=E'(\mu)=\frac{x}{L+1}F'(x)= \frac{A\alpha}{L+1} x^{\alpha}+\mathcal{O}(x^{\alpha+1}),
\end{equation}
so that
\begin{align}
x&= \biggl(\frac{(L+1)j}{A\alpha}\biggr)^{1/\alpha}+\mathcal{O}(j^{1/\alpha+1})\\
\mu&=\frac{L+1}{\alpha}\log\biggl(\frac{(L+1)j}{A\alpha}\biggr)+\mathcal{O}(j)\\
E(\mu)&=-z_0+\frac{(L+1)j}{A\alpha}+\mathcal{O}(j^2)
\end{align}
yielding finally
\begin{equation}
g(j)=z_0+ \frac{(L+1)}{\alpha}\Biggl(j\log\biggl(\frac{(L+1)j}{A\alpha}\biggr)-j\Biggr)+\mathcal{O}(j^2)
\end{equation}
so that the Legendre transform of $-z_0+A{\rm e}^{\alpha\mu/(L+1)}\sim E(\mu)$ is indeed an equivalent of $g(j)$.

\section{Identity on escape rates for two-body interactions}
\label{A2}

In this appendix, we prove the claim made in section \ref{IVd1} that by choosing $a_k=k$ for $k\leq K$ and $a_k=1$ otherwise, all the escape rates of anti-shocks of width $K$ or less are equal to $1$, and all escape rates from other states are larger.

Consider an integer $n$, complex numbers $\{a_i\}_{i:1..n}$ and $\{b_j\}_{j:1..n+1}$, and the complex function
\begin{equation}
f(z)=\frac{\prod\limits_{i=1}^{n}(a_i-z)}{\prod\limits_{j=1}^{n+1}(b_j-z)}.
\end{equation}

That function being rational, the sum of all its residues is $0$. It has poles at all the $b_j$'s and at infinity, and the residue at infinity is clearly $-1$, which yields:
\begin{equation}
\sum\limits_{l=1}^{n+1}\frac{\prod\limits_{i=1}^{n}(a_i-b_l)}{\prod\limits_{j=1,j\neq l}^{n+1}(b_j-b_l)}=1.
\end{equation}

Consider now $0=b_1<a_1<b_2<a_2<...<b_n<a_n<b_{n+1}\leq K$ to be integers, corresponding to the positions of alternating $10$ and $01$ boundaries in an anti-shock state of width $K$ or less, as seen on fig.\ref{fig-LongStates}. Every term in the sum then corresponds precisely to the jump rate of the particle at position $b_l$, so that the whole sum is the escape rate from that configuration, equal to $1$. Moreover, the escape rate from an anti-shock with a width larger than $K$ can be obtained from that expression by replacing every term $(a_i-b_l)$ or $(b_j-b_l)$ which is larger than $K$ in absolute value by $\pm1$, making the corresponding ratio, and hence the whole sum, strictly larger.

\section{Left-right duality for the biased ASEP}
\label{A3}

In this appendix, we derive a surprising symmetry for the large deviation function of the standard one-dimensional open ASEP. Consider once more the deformed Markov matrix
\begin{equation}
M_{\mu}=m_0(\mu_0)+\sum_{i=1}^{L-1} M_{i}(\mu_i)+m_L(\mu_l)
\end{equation}
with
\begin{equation}
m_0(\mu_0)=\begin{bmatrix} -p_0 & q_0{\rm e}^{-\mu_0} \\ p_0{\rm e}^{\mu_0} & -q_0  \end{bmatrix}~,~ M_{i}(\mu_i)=\begin{bmatrix} 0 & 0 & 0 & 0 \\ 0 & -q & p{\rm e}^{\mu_i} & 0 \\ 0 & q{\rm e}^{-\mu_i}& -p & 0 \\ 0 & 0 & 0 & 0 \end{bmatrix}~,~m_L(\mu_L)=\begin{bmatrix} -q_L & p_L{\rm e}^{\mu_L} \\  q_L{\rm e}^{-\mu_L} & -p_L  \end{bmatrix}.
\end{equation}
It was noticed in \cite{Torkaman2015} that for certain constrained boundary rates, which can be written as $p_0=p\rho_a $, $q_0=q(1-\rho_a)$, $p_L=p(1-\rho_b)$ and $q_L=q\rho_b $, the deformed Markov matrix for $\mu=\log(q/p)$ was, up to a constant, the same as the non-deformed Markov matrix with $p\leftrightarrow q$ and $\rho_{a,b}\leftrightarrow(1-\rho_{a,b})$, which can be seen as either a left/right or a particle/hole transformation. We will see here that this identity can be extended to a symmetry of the whole deformed Markov matrix, and by extension of the cumulant generating function $E(\mu)$ and the large deviation function $g(j)$. In order to do that, we will need to add $p$, $q$, $\rho_a$ and $\rho_b$ as parameters to our notations, i.e. for instance write the deformed Markov matrix as $M_{\mu}^{p,q;\rho_a,\rho_b}$.

We now consider the following identities (remembering that we note the occupancy of site $i$ as $\tau_i$):
\begin{align}
m_0^{p,q;\rho_a}\bigl(\log(q/p)+\mu_0\bigr)&=m_0^{q,p;\rho_a}\bigl(\mu_0\bigr)-(p-q)\rho_a+(p-q)\delta_{\tau_1,1}\\
M_{i}^{p,q}\bigl(\log(q/p)+\mu_i\bigr)&=M_{i}^{q,p}\bigl(\mu_i\bigr)+(p-q)(\delta_{\tau_{i+1},1}-\delta_{\tau_i,1})\\
m_L^{p,q;\rho_b}\bigl(\log(q/p)+\mu_L\bigr)&=m_L^{q,p;\rho_b}\bigl(\mu_L\bigr)+(p-q)\rho_b-(p-q)\delta_{\tau_L,1}.
\end{align}
Summing those identities, we see that all the deltas cancel one-another, and we are left with
\begin{equation}
M_{\mu'}^{p,q;\rho_a,\rho_b}=M_{\mu}^{q,p;\rho_a,\rho_b}+(p-q)(\rho_a-\rho_b)
\end{equation}
with $\mu'=(L+1)\log(q/p)+\mu$ (remember that $\mu=\sum_i \mu_i$), which leads to the same identity for $E^{p,q;\rho_a,\rho_b}(\mu)$:
\begin{equation}
E^{p,q;\rho_a,\rho_b}\bigl((L+1)\log(q/p)+\mu\bigr)=E^{q,p;\rho_a,\rho_b}(\mu)+(p-q)(\rho_b-\rho_a).
\end{equation}
Moreover, we have that
\begin{align}
E^{p,q;\rho_a,\rho_b}(\mu)&=E^{q,p;\rho_b,\rho_a}(-\mu)\\
&=E^{q,p;1-\rho_a,1-\rho_b}(-\mu).
\end{align}
The first equality is obtained through a left/right transformation $\tau_i\leftrightarrow\tau_{L+1-i}$, and the second through a particle/hole transformation $\tau_i\leftrightarrow 1-\tau_i$. In both cases, the current is reversed, hence the $-\mu$. Finally, the Gallavotti-Cohen symmetry reads
\begin{equation}
E^{p,q;\rho_a,\rho_b}(\mu)=E^{p,q;\rho_a,\rho_b}\biggl(\log\Bigl(\frac{q^{L+1}(1-\rho_a)\rho_b}{p^{L+1}\rho_a(1-\rho_b)}\Bigr)-\mu\biggr).
\end{equation}
To see the consequence of those symmetries on the dynamical phase diagram (as shown on fig.\ref{fig-PhaseSigma}.a), we have to take the $L\rightarrow\infty$ limit, which leads to two simplifications: first, the part of the phase diagram corresponding to $j\leq 0$, i.e. $\mu\leq \frac{1}{2}\log\Bigl(\frac{q^{L+1}(1-\rho_a)\rho_b}{p^{L+1}\rho_a(1-\rho_b)}\Bigr)$, is rejected to $\mu\rightarrow-\infty$ and disappears from the diagram ; secondly, as is remarked in section IV.C of \cite{Lazarescu2015}, the value of $E(\mu)$ does not depend on all four of the boundary rates, but only on two combinations which are here precisely equal to $\rho_a$ and $\rho_b$, meaning that the special case we considered is generic in the large size limit. We can therefore describe the symmetries of the full phase diagram by combining all the finite-size relations that we have just written, keeping only those for the forward ASEP ($p,q$ in that order, which we can remove from the parameters) and for $\mu$ finite. The left/right $+$ particle/hole symmetry $E^{\rho_a,\rho_b}(\mu)=E^{1-\rho_b,1-\rho_a}(\mu)$ is well known and clearly visible on the diagram, but we also get a new and surprising symmetry which exchanges the two reservoirs:
\begin{equation}
\boxed{
E^{\rho_a,\rho_b}\biggl(\mu+\log\Bigl(\frac{(1-\rho_a)\rho_b}{\rho_a(1-\rho_b)}\Bigr)\biggr)=E^{\rho_b,\rho_a}(\mu)+(p-q)(\rho_b-\rho_a).
}
\end{equation}
This symmetry exchanges the shock and anti-shock phases as seen on fig.20 of \cite{Lazarescu2015}, and moreover, since all those calculations can be done in much the same way directly on the deformed Markov matrices (which we did not do here simply for the sake of readability), we find that the typical profiles associated to shocks and anti-shocks are symmetric to one-another. In particular, the phase transitions shown in dark purple on fig.\ref{fig-PhaseSigma}.a are equivalent.

\newpage

\bibliographystyle{mybibstyle}

\bibliography{Biblio}{}

\end{document}